\documentclass[12pt]{article}
\usepackage{a4wide}
\usepackage[dvipdfmx,hiresbb]{graphicx} 
\usepackage{mediabb}                    

\usepackage{color}
\usepackage{slashed}

\newcommand{\al}[1]{\begin{align}#1\end{align}}

\newcommand{\paren}[1]{\left(#1\right)}

\newcommand{\sqbr}[1]{\left[#1\right]}

\newcommand{\vev}[1]{\left\langle#1\right\rangle}

\newcommand{\GeV}{\ensuremath{\,\text{GeV} }}

\newcommand{\nn}{\nonumber\\}

\usepackage{amsmath,amssymb}
\usepackage{epsf}

\begin{document}
\title{\vbox{
\baselineskip 14pt
\hfill \hbox{\normalsize KEK-TH-1938}\\
} \vskip 1cm
\bf \Large 
Baryon asymmetry from primordial black holes
\vskip 0.5cm
}
\author{
Yuta~Hamada$^{1,2}$\thanks{E-mail: \tt yhamada@wisc.edu},~
Satoshi~Iso$^{1,3}$\thanks{E-mail: \tt satoshi.iso@kek.jp},
\\*[20pt]
{\it \normalsize
$^1$KEK Theory Center, IPNS, KEK, Tsukuba, Ibaraki 305-0801, Japan} \\
{\it \normalsize
$^2$Department of Physics, University of Wisconsin, Madison, WI 53706, USA} \\
{\it \normalsize $^3$ Graduate University for Advanced Studies (SOKENDAI),  305-0801, Japan} 
\smallskip
} 
\date{\today}


\maketitle

\abstract{\noindent \normalsize
We propose a new scenario of the baryogenesis from  primordial black holes (PBH). 
Assuming presence of microscopic baryon (or lepton) number violation, and presence of an effective 
CP violating operator such as $\partial_\alpha F(\mathcal{R_{....}} ) J^\alpha$, where 
$F(\mathcal{R_{....}})$ is a scalar function of the Riemann tensor and $J^\alpha$ is a baryonic (leptonic) current,  time evolution of an evaporating black hole 
 generates  baryonic (leptonic) chemical potential at the horizon; 
consequently  PBH emanates  asymmetric Hawking radiation between baryons (leptons) and anti-baryons (leptons).
Though the operator is higher dimensional and largely suppressed by a high mass scale $M_*$, 
we show that 
 sufficient amount of asymmetry can be generated for a wide range of parameters of the PBH mass $M_{\rm PBH}$, its abundance $\Omega_{\rm PBH}$, 
 and the
scale $M_*$.  
}

\newpage
\normalsize
\section{Introduction}
The standard model of particle physics is completed by the discovery of the Higgs boson, and is
surprisingly consistent with the experimental data up to $1$~TeV scale.
However, there still remain several unsolved questions, e.g., 
What is the dark matter in the universe?
Why are  baryons more abundant than  anti-baryons?

In order to answer these questions,  gravitational effects might play important roles.
One of the interesting possibilities of  gravitational effects will be  
primordial black holes (PBH)~\cite{Carr:1974nx,Carr:1975qj}, which may be created  in the early universe. 
PBHs could be formulated in the early universe by various processes such as large density fluctuations by inflation~\cite{GarciaBellido:1996qt}, preheating~\cite{Taruya:1998cz},
in particular the tachyonic preheating~\cite{Felder:2000hj},  
or bubble collisions~\cite{Crawford:1982yz} associated with  first order phase transitions in the universe~\cite{Hawking:1982ga}.\footnote{
See also Ref.~\cite{Kuhnel:2015qaa} for discussion on PBH formation within the recently proposed framework of graviton condensates~\cite{Dvali:2011aa}.
} 
A PBH evaporates by  Hawking radiation until the present time if its mass $M$ is lighter than $M=10^{15}$g.
Consequently the abundance of the PBHs around $M=10^{15}$ g is strongly constrained by 
observations of the cosmic gamma ray \cite{MacGibbon:1991vc,Carr:2016hva}. If the mass is between $10^9$ g  and $10^{13}$ g,
the Hawking radiation from the PBHs affect the big bang nucleosynthesis (BBN) and the abundance in this mass region 
is also strongly constrained (see e.g.~\cite{Carr:2009jm}).   
PBHs with larger mass also play various important roles in cosmology.
It may contribute  to the dark matter in the universe for $M\gtrsim10^{15}$g although its abundance is severely constrained~\cite{Carr:2016drx,Chen:2016pud}.
PBH may also explain the origin of the BHs with mass $M=\mathcal{O}(10^{30})$g~\cite{Bird:2016dcv,Sasaki:2016jop}, whose binary mergers are 
observed in the recent detections of the gravitational waves  by LIGO~\cite{Abbott:2016blz}.

On the other hand, PBHs with smaller mass $M<10^{8}$ g will play a different role. 
One of the  important roles of lighter PBHs will  be to  give a stage for  generating baryon asymmetry.
Hawking~\cite{Hawking:1974rv}, Carr~\cite{Carr:1976zz} and Barrow~\cite{Barrow01111979}
 proposed a scenario of Baryogenesis in which 
GUT scale particle/right handed neutrino  are created by the Hawking radiation
and then decay in a C and CP violating manner (see also \cite{Barrow:1981zv,Barrow:1990he,Baumann:2007yr,Fujita:2014hha}). 
Recently Hook proposed a different scenario of Baryogenesis by using asymmetric Hawking radiation due to 
a dynamically generated baryonic chemical potential at the horizon \cite{Hook:2014mla}.
There the CP violating interaction of the baryonic (or leptonic) current $J^\alpha$ and the scalar curvature $\mathcal{R} $
 \al{ \label{simplestCPviolating-op}
{1\over M_*^2}\partial_\alpha 
\mathcal{R} 
J^\alpha
,
}
is assumed, and the time evolution of the universe is used to generate the chemical potential $\mu=\dot{\mathcal R}/M_*^2$ for baryons.
The same interaction is used in the gravitational baryogenesis~\cite{Davoudiasl:2004gf}. 
Indeed, the mechanism \cite{Hook:2014mla} essentially 
utilizes the idea of the spontaneous baryogenesis~\cite{Cohen:1987vi,Cohen:1988kt} and gravitational baryogenesis~\cite{Davoudiasl:2004gf}  scenarios.

In this paper, we propose a new mechanism of the baryogenesis from evaporating PBHs.
The mechanism is similar to that of \cite{Hook:2014mla},  
but instead of using the time evolution of the universe, 
 we make use of the time evolution of the mass of the PBH itself for generating the Baryonic chemical potential\footnote{ \label{footnote2}
 A similar idea is proposed in  \cite{Banks:2015xsa}. 
 Note that, by chemical potential,  we here mean asymmetry of propagations between particles and anti-particles due to the interaction with the background geometry.
 If particles enter thermal equilibrium, the distributions become asymmetric.
 In the case of Hawking radiation, the radiation from black holes becomes asymmetric as if there is a chemical potential. 
 Hence, in the present paper we call the term $\mu J^0$ a chemical potential term.
 }. 
 This leads to a big difference between our mechanism and  \cite{Hook:2014mla}.
 Since the scalar curvature around the PBH in  vacuum is vanishing, we need to  
use higher dimensional operators such as 
\al{\label{Eq:operator}
{1\over M_*^4}\partial_\alpha 
\paren{\mathcal{R}_{\mu\nu\rho\sigma}\mathcal{R}^{\mu\nu\rho\sigma}} 
J^\alpha
,
}
where $\mathcal{R}_{\mu\nu\rho\sigma}$ is the Riemann tensor\footnote{ \label{footnote:Gauss} 
Our investigation  does not depend much on the specific
form of the higher dimensional operators. 
We can instead use  the Gauss-Bonne type 
\begin{equation}
\partial_\mu(\mathcal{R}^2+\mathcal{R}^{\mu\nu\rho\sigma}\mathcal{R}_{\mu\nu\rho\sigma}-4\mathcal{R}_{\mu\nu}\mathcal{R}^{\mu\nu})J^\mu.
\end{equation}
which can be transformed to the scalaron picture~\cite{Kobayashi:2011nu}, and is safe in view of the ghost modes.
The first and the third terms in the Gauss-Bonne term vanish around the Schwartzschild BH, so it give the same
effective CP-violating operator as the one we introduced.
}. The operator is largely
 suppressed by the scale $M_*$
and seems to be negligible but we show that it is sufficient to generate the desired asymmetry.
It is due to the fact that 
 the dynamically generated chemical potential $\mu$, as well as 
the temperature of the Hawking radiation $T_H$, is time-dependent through the mass of the BH. 
Indeed  $\mu/T_H$ is increasing as the PBH evaporates and becomes larger than 1
at the late stage of the evaporation. 
Consequently the asymmetry in the Hawking radiation becomes maximal after the PBH mass becomes smaller 
than a critical mass. The critical mass is determined by  the scale $M_*$. 

Once the chemical potential is generated at the horizon, Hawking radiation  produces lepton/baryon asymmetry.
Since the sphaleron process~\cite{Manton:1983nd,Klinkhamer:1984di,Kuzmin:1985mm}  violates $B+L$, it is necessary to generate $B-L$ 
if the typical Hawking temperature is higher than 100 GeV. 
We thus assume violation of the $B-L$ number in the underlining microscopic theories 
such as  interactions with right-handed neutrinos 
or some effects related to the quantum gravity ~\cite{Giddings:1987cg}.
The CP symmetry is broken by the effective operator  (\ref{Eq:operator}).
The time dependence of the PBH mass due to Hawking radiation induces time dependent and position dependent chemical potential,  which is apparently a non-equilibrium process. 
In this way, Sakharov's three conditions~\cite{Sakharov:1967dj} are satisfied in the present model\footnote{It is often stated that the Sakharov's three conditions are not necessary in the spontaneous baryogenesis.
In the present scenario, since the CP parity of $F({\cal R....})\propto {\cal R_{\mu \nu \lambda \sigma}}^2$ is even, the higher dimensional operator  breaks C and CP symmetries. }.

The paper is organized as follows.
In the next section, we explain the basic mechanism of the scenario,  and estimate the order of the asymmetry.
We show that in some region of the parameter space of the PBH mass $M_{\rm PBH}$ and the scale $M_*$,
the desired asymmetry 
$n_B/s\simeq8.7\times10^{-11}$~\cite{Ade:2015xua} can be generated.
A possible origin of the higher dimensional operator  (\ref{Eq:operator}) is given 
in Sec.~\ref{Sec:higher_dimensional_operators}, and  we 
 estimate the order of the scale $M_*$.
In Sec.~\ref{Sec:washout} we show that washout of the generated lepton number outside the horizon
does not occur for the typical interaction discussed in Sec.~\ref{Sec:higher_dimensional_operators}.
Sec.~\ref{Sec:conclusion} is devoted to the summary and discussions.
In Appendix A and B, we discuss Hawking radiation with the chemical potential. In Appendix C, we give an analytical approximation
of the function  $g_n(X)$ used in Sec.\ref{Sec:main}. 
\section{Baryo (lepto)-genesis at the BH horizon}  \label{Sec:main}
\subsection{CP-violating interactions} \label{section-2.1}
The scenario of the gravitational baryogenesis \cite{Davoudiasl:2004gf} assumes the CP-violating interaction (\ref{simplestCPviolating-op}) where
$M_*$ is the scale of the underlying theory that generates such an interaction.
In an expanding universe, the time derivative of the scalar curvature $\dot{\mathcal{R}}$ is 
non-zero and the interaction  generates a chemical potential\footnote{
As explained in footnote \ref{footnote2}, 
 the energy spectrum becomes asymmetric  between particles and anti-particles.
Then,  as long as typical time scale of the interaction is smaller than that of the expansion of the universe,  particles enter in thermal equilibrium and
$ \mu J^0 $  term can be interpreted as a chemical potential.  } 
$\mu=\dot{\mathcal{R}}/M_*^2$.
If a $B$-violating interaction is present\footnote{In the present section \ref{section-2.1}, for simplicity, we use B(aryon) to represent
the current $J^\mu$. It can be either baryons or leptons but a necessary condition is that it has non-vanishing $B-L$ charge.} and the system is in thermal equilibrium, the  distribution becomes asymmetric between baryons and anti-baryons.
Then once the temperature drops below the freezing-out temperature of the B-violating interaction, the asymmetry remains in the later universe. 
The scenario is applied to the evaporating BH by the Hawking radiation \cite{Hook:2014mla}. 
The term $\mu J^0$ is similarly generated by the evolution of the universe
and the Hawking radiation becomes asymmetric, but the freezing-out scenario is different.
Since the thermal radiation from black holes is created, not by  thermal process of B-violating interaction,
but by  genuine quantum process, the condition of the thermal equilibrium and the 
freezing-out  in \cite{Davoudiasl:2004gf}  is not necessary to be introduced in the analysis of \cite{Hook:2014mla}. 

In this section, we generalize the idea of the gravitational baryogenesis from  a PBH \cite{Hook:2014mla}
by taking the direct effect of the decay of the PBH mass $M(t)$. Since the scalar curvature $\mathcal{R}$ vanishes
outside of the BH in  vacuum\footnote{In \cite{Davoudiasl:2004gf} and \cite{Hook:2014mla},  
the radiation dominated universe with the trace anomaly of the energy momentum tensor is studied 
so as to make $\mathcal{R}$ non-vanishing.}, we  consider an operator such as in (\ref{Eq:operator}). 
More generally, we can consider a class of higher dimensional operators\footnote{
In addition to the CP-violating interaction, 
we implicitly assume that the existence of  B-violating operator since otherwise this term vanishes by performing an integration by parts.
See Sec.~\ref{Sec:higher_dimensional_operators} and App.~\ref{App:chemical potential} for the discussion of  the physical meaning of this operator.
}
,
\al{\label{Eq:operator2}
&
{a_n\over M_*^{4n}}\partial_\alpha 
\paren{\mathcal{R}_{\mu\nu\rho\sigma} \mathcal{R}^{\mu\nu\rho\sigma}
}^n 
J^\alpha
.
&
n\geq2.
}
It can be further generalized to
\al{ \label{mostgeneralop}
\partial_\alpha F(\mathcal{R_{....}}) J^\alpha
}
where $F(\mathcal{R_{....}})$ is any scalar function made of the curvature tensors.
For the square of the Riemann curvature of the Schwartzschild BH is given by\footnote{See, e.g., Ref.~\cite{gravitation}.}
\al{\label{Eq:Riemann_square0}
\mathcal{R}_{\mu\nu\rho\sigma}\mathcal{R}^{\mu\nu\rho\sigma}
={3M^2\over4\pi^2 M_P^4 r^6}
,
}
a non-vanishing chemical potential $\mu=a_n \partial_0 \paren{\mathcal{R}_{\mu\nu\rho\sigma}\mathcal{R}^{\mu\nu\rho\sigma}}^n /M_*^{4n}$
is generated if the BH mass $M$ is decaying. 
Here we have introduced the reduced Planck scale,
\al{
M_P:=\paren{8\pi G}^{-1/2}=2.43\times10^{18}\GeV=4.3\times10^{-6}g=\paren{2.7\times10^{-43}s}^{-1},
}
where $G$ is the Newton constant.
Note that the chemical potential is dependent on  time through $M(t)$. It also changes  with the distance $r$   from the BH. 
Since the Hawking radiation is generated by the Bogoliubov transformation between the vacua of quantum fields
near the horizon and at  far-infinity from the BH, 
the chemical potential near the horizon is relevant to generate the asymmetry of the Hawking radiation. 
The propagations of baryon and anti-baryon become different in the vicinity of the horizon, which shift the energy between them.
Accordingly the generated asymmetry is proportional to the chemical potential evaluated at the horizon $r=r_H$. 
Here we note that even if we instead evaluate the chemical potential at, e.g., $r \simeq 2 r_H$, it does not change our conclusion very much
(see the 2nd paragraph of Sec. \ref{Sec:conclusion}).

\subsection{Basic properties of evaporating BH}
We  summarize some basic facts about an evaporating BH. For simplicity\footnote{In general, BH can have charge and angular momentum, but these would be quickly lost before most of BH mass disappears~\cite{Carter:1974yx,Gibbons:1975kk,Page:1976ki}.} we consider the Schwarzschild black hole
with the metric, 
\al{\label{Eq:Schwarzschild}
ds^2
=
\paren{1-{2GM\over r}}dt^2-{1\over \paren{1-\dfrac{2GM}{r}}}dr^2- r^2 d\theta^2 - r^2 \sin^2\theta d\varphi^2,
}
where $M$ is the mass of the black hole.
The radius of the horizon and Hawking temperature are given by
\al{
r_H&=2GM
={M\over4\pi M_P^2},
\nn
T_H&={M_P^2\over M}.
}

Through the Hawking radiation,  particles are emitted ~\cite{Hawking:1974sw} from the BH with the rate
\al{
{dN\over d\omega dt}
={1\over2\pi}{\Gamma\over e^{\omega/T_H}\pm1},
\nn
{dE\over d\omega dt}
={1\over2\pi}{\Gamma\omega\over e^{\omega/T_H}\pm1},
}
where $N, E$ are  number and energy of emitted particles, $\omega$ is the frequency, and $\Gamma$ is the absorption probability (or gray body factor), 
which is caused by gravitational scatterings of emitted particles outside  the horizon. The absorption probability depends on particle species, especially
on the spin of the emitted particles. 
At low frequency, $\omega\to0$, the absorption cross sections $\sigma:=\pi\omega^{-2}\Gamma$ of spin $0$ and $1/2$ particles are constant while those of spin $1$ and $2$ particles are 
proportional to square and fourth power of frequency, respectively:
\al{
\sigma&\to 	\begin{cases}
			\text{const.}	&	
				\text{for spin $0$ and $1/2$,}\\ \\
			\omega^2	&	
				\text{for spin $1$,}\\ \\
			\omega^4 &	
				\text{for spin $2$.}		
		\end{cases}
.		
}
%
As a result, most of the energy emitted from the PBH is carried by scalars and fermions~\cite{Page:1976df}.

We emphasize that the spectrum of the Hawking radiation is (almost) thermal, not because  thermal plasma
at temperature  $T_H$ is realized due to sufficiently fast interactions between emitted particles, but simply because 
the quantum vacuum at the horizon behaves as if it is in the thermal equilibrium for an observer at  far-infinity. 
In fact, even extremely weakly coupled particles (such as  gravitons), that can be thermalized only at $T_H \gtrsim M_P$,
are emitted according to the Hawking thermal spectrum.

Once we take into account the Hawking radiation, the spacetime is no longer stationary, and the metric~\eqref{Eq:Schwarzschild} is no longer appropriate to describe the evaporating BH.
Since the mass in~\eqref{Eq:Schwarzschild}  is the ADM mass which includes the energy of the emitted radiation 
it cannot correctly describe the mass of a decaying BH itself. 
A simplest alternative is the outgoing Vaidya metric~\cite{Vaidya:1951zz}, which is a solution 
of the Einstein equation describing outgoing null dust:
\al{\label{Eq:outgoing Vaidya}
ds^2
=
\paren{1-{2GM(u)\over r}}du^2+2dudr- r^2 d\theta^2 - r^2 \sin^2\theta d\varphi^2,
}
where $u=t-r_*, r_*=r+2M\log|(r-2M)/2M|$.
The apparent horizon is located at $r=r_H=2GM(u)$.
The corresponding energy momentum tensor is given by
\al{&
T_{\mu\nu}
=
-{dM\over du}{1\over4\pi r^2}l_\mu l_\nu,
&&
l_\mu=\partial_\mu u,
}
which describes dust with energy density $\rho=(-dM/du)/(4\pi r^2)$ moving with a four-velocity $l^\mu$.
The mass $M(u)$ represents the Bondi mass, which is nothing but the mass of the BH itself. 
In the following, we consider the time-evolution of the Bondi mass $\partial_u M(u)$.

Because of the (almost) thermal Hawking radiation, 
the black hole loses its energy following
\al{\label{Eq:mass_change}
{dM\over d u}
\simeq
-\paren{8\pi}^2{M_P^4\over M^2}\alpha,
}
where $\alpha$ is a numerical coefficient~\cite{Page:1976df} which can be determined by taking the 
effects of absorption cross section $\sigma.$
As we discussed above, 
 the dominant contribution to $\alpha$ in the standard model comes from fermions. 
In Ref.~\cite{Page:1976df}, it is shown that the contribution to $\alpha$ from $\nu_e$ and $\nu_\mu$ is
$1.575\times10^{-4}$.
Then, summing up all the fermionic degrees of freedom in the standard model, we obtain\footnote{
Here we assume that the value of $\alpha$ is the same to all the fermonic degrees of freedom,
and the coefficients in the parenthesis are 1 for the $SU(2)_L$ singlet lepton, 2 for the doublet lepton,
and  $4 \times 3$ for up and down quarks with color degrees of freedom. 
The coefficient  $3/2$ is a transformation factor from the 2 generation calculation to 3.
}
\al{
\alpha
&=1.575\times10^{-4}
\times{3\over2}\times\paren{1+2+4\times3}
\nn
&=3.5\times10^{-3}.
}
%
Solving the Eq.~\eqref{Eq:mass_change},  the time dependence of $M$ is given by
\al{\label{Eq:lifetime}&
M(u)=
M_\text{PBH}
\paren{
1-{u-u_\text{ini}\over \tau}
}^{1/3},
}
and the lifetime of the black hole $\tau$ becomes
\al{
\tau=
{M_\text{PBH}^3\over M_P^4}{1\over3\paren{8\pi}^2 \alpha}.
}
Here we take $M=M_\text{PBH}$ at the initial time $u=u_\text{ini}$.
From Eq.~\eqref{Eq:lifetime}, we can see that the PBH completely evaporates until today if $M_\text{PBH}$ is smaller than $10^{20}M_P\sim10^{15}g$.

\subsection{Dynamically generated chemical potential}
Since the square of the Riemann tensor  outside  a PBH is 
\al{\label{Eq:Riemann_square}
\mathcal{R}_{\mu\nu\rho\sigma}\mathcal{R}^{\mu\nu\rho\sigma}
={12r_H^2\over r^6}
={3M(u)^2\over4\pi^2 M_P^4 r^6}
,
}
we have a chemical potential $\mu$, if the CP-violating interaction Eq.~\eqref{Eq:operator} is present,
\al{\label{Eq:chemical_potential1}
\mu=
{3\over2\pi^2}{M\over M_P^4 M_*^4 r^6}{dM\over du}
\simeq
-96\alpha{1\over M M_*^4 r^6}
.
}
By taking $r=r_H$, the chemical potential evaluated at the horizon becomes
\al{\label{Eq:chemical_potential2}
\mu|_{r=r_H}&=
-{3\over2}(8\pi)^6\alpha
M_P \paren{M_P\over M}^7\paren{M_P\over M_*}^4.
}
Then the ratio of $\mu |_{r_H}$ to the Hawking temperature is $T_H=M_P^2/M$ is given by
\al{\label{ratio-mu/T}
\frac{\mu |_{r_H}}{T_H} = -
{3\over2}(8\pi)^6\alpha
 \paren{M_P\over M}^6 \paren{M_P\over M_*}^4 
 = - \paren{M_{\rm cr} \over M}^6,
}
where we have defined the critical mass $M_{\rm cr}$ by
\al{ \label{criticalmass}
M_{\rm cr} = 8 \pi M_P \left( \frac{3 \alpha}{2} \right)^{1/6} \left( \frac{M_P}{M_*} \right)^{2/3} \sim 10 \times \left( \frac{M_P}{M_*} \right)^{2/3}  M_P
.
}
Note that the (absolute value of the) ratio is  increasing as the BH mass $M$ decreases.
For $M < M_{\rm cr}$,  the ratio exceeds 1 and the asymmetry of the radiation becomes maximal.
It indicates that if the initial mass of the PBH is smaller than the critical mass, only baryons are emitted \footnote{Of course, particles
without the baryon number are also emitted.}.

Parametrizing  the scale $M_*$ as $M_*=10^x M_P$ and  the initial mass of the PBH as $M_{\rm PBH}=10^y M_P$,
the condition of $M_{\rm PBH} = M_{\rm cr}$ becomes
\al{
y = 1 -{2 \over 3} x .
}
We will see later that the relation plays an important role in generating the asymmetry.
\subsection{Leptogenesis from a PBH}

Now let us explicitly calculate the asymmetry produced by the evaporation of $one$ PBH. 
Since we have in mind a model in which the CP-violating interaction is induced by 
the interactions with the right-handed neutrinos,  we hereafter suppose that the
chemical potential induced at the horizon is  the leptonic one. 
Then, as we see that the temperature of the universe at the epoch of evaporation is much higher than the 
electroweak scale, sphaleron processes transmute the generated leptons into baryons. 

First note that the averaged energy per each emitted massless particle is
\al{
\vev{E}
=
{n_{\rm other}\over n_{\rm tot}}\vev{E_{\rm other}}+{n_L\over n_{\rm tot}}\vev{E_L}+{{n_{\bar L}}\over n_{\rm tot}}\vev{E_{\bar L}},
}
where the subscripts $L(\bar{L})$ and { \it other} represent the leptons (anti-leptons) and the other emitted particles (including  only  scalars and fermions) in the standard model respectively.
 $n_i$ is the number density for each species and given by
 \al{
&
n_{\rm other}
={g_{\rm other}\over(2\pi)^3}\int  d^3 k \  \paren{e^{k/T_H}+1}^{-1}
=g_{\rm other} {3\zeta(3)\over4\pi^2} T_H^3,
\nn
&
n_L
={g_L\over(2\pi)^3}\int  d^3 k \  \paren{e^{(k+\mu)/T_H}+1}^{-1}
=-{g_L\over\pi^2}T_H^3\rm{Li}_3 \paren{-e^{-\mu/T_H}},
\nn
&
n_{\bar L}
=-{g_{\bar L}\over\pi^2}T_H^3\rm{Li}_3 \paren{-e^{\mu/T_H}}, 
}
where $g_i$ is the internal degrees of freedom.
In the case of the standard model, $g_L=g_{\bar{L}}=9$ and $g_{\rm other}=76$.
$n_{\rm tot}=n_{\rm other}+n_L+n_{\bar L}$ is the total number. 
${\rm Li}_a (z)$ is the  polylogarithmic function  defined by ${\rm Li}_a(z):=\sum_{k=1}^\infty z^k/k^a$.
In the present convention, $n_L > n_{\bar L}$ since $\mu<0$. It is, of course, reversed if the sign of the coefficient of the  CP violating operator is reversed.
 $\vev{E_i}$ is the averaged energy for each species 
and the explicit expressions are given by
\al{
&
\vev{E_{\rm other}}
={\int  d^3 k \  k \paren{e^{k/T_H}+1}^{-1} \over \int  d^3 k \paren{e^{k/T_H}+1}^{-1} }
={7\pi^4\over180 \zeta(3)}T_H,
\nn
&
\vev{E_L}
={\int  d^3 k \  k \paren{e^{(k+\mu)/T_H}+1}^{-1} \over \int  d^3 k \paren{e^{(k+\mu)/T_H}+1}^{-1} }
=3T_H{\rm{Li}_4 \paren{-e^{-\mu/T_H}}\over \rm{Li}_3 \paren{-e^{-\mu/T_H}}},
\nn
&
\vev{E_{\bar L}}
=3T_H{\rm{Li}_4 \paren{-e^{\mu/T_H}}\over \rm{Li}_3 \paren{-e^{\mu/T_H}}}.
}
Notice that $\vev{E_L}$ behaves as $\vev{E_L} \sim T_H$
 for $|\mu|<T_H$, and $\vev{E_L}  \sim |\mu|$ for $T_H<|\mu|$.

\begin{figure}
\begin{center}
\hfill
\includegraphics[width=.6\textwidth]{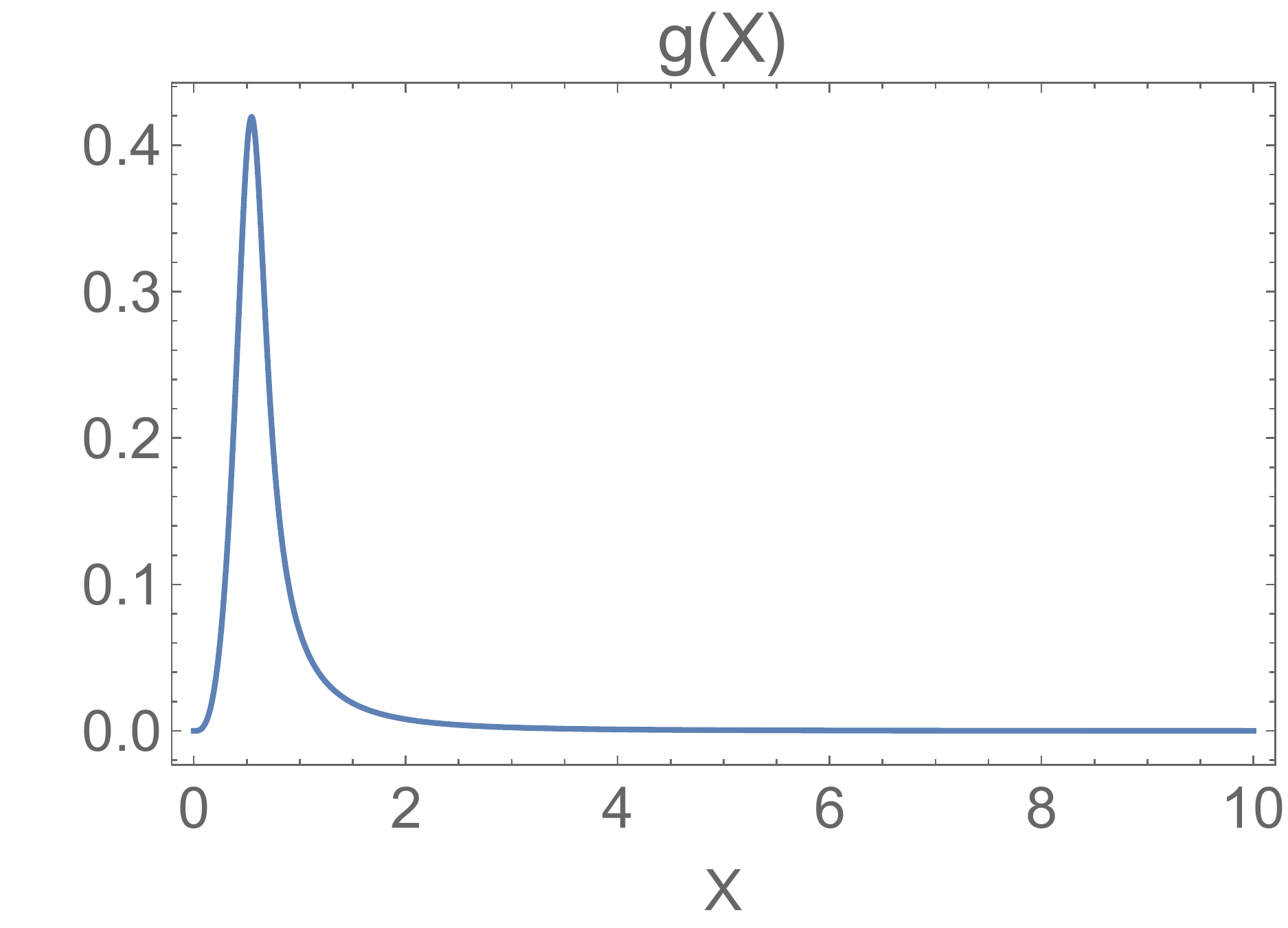}
\hfill\mbox{}
\end{center}
\caption{
The function $g(X)$ is plotted with the SM values, $g_L=g_{\bar{L}}=9$ and $g_{\rm other}=76$.
Since the generated lepton number is proportional to the integral of $g(X)$,
the asymmetry is efficiently produced around 
$X \sim 0.5$, namely $M \sim 0.7 M_{\rm cr}.$ 
For larger values of $X=(M/M_{\rm cr})^2$, the chemical potential $|\mu|$ becomes negligibly small. For smaller values, the asymmetry is maximal
but the number of emitted particles is reduced due to large $|\mu|$. 
}
\label{Fig:integral}
\end{figure}

By using these formula,  the lepton number asymmetry can be estimated as
\al{ \label{Eq:delta NL}
\delta N_L
&=\int_{M_{\rm min}}^{M_\text{PBH}} {dM\over \vev{E}}{n_L+n_{\bar L}\over n_{\rm tot}}{n_L-n_{\bar L}\over n_L+n_{\bar L}}.
}
Here we introduced the lower cutoff 
$M_{\rm min}$ for the mass of the PBH, under which the typical energy scale of the Hawking radiation becomes higher than $M_*$ and the 
present analysis becomes questionable.
It is determined by the condition, either $T_H=M_*$ or $|\mu|=M_*$, and given by
\al{
M_{\rm min}=
{\rm max}
\paren{ \frac{M_P^2}{ M_* }  ,   \paren{M_{\rm cr}^6 M_P^2\over M_*}^{1/7}     }.
}
For $M<M_{\rm min}$, either $T_H$ or $|\mu|$ is larger than the scale $M_*$, and 
 the present investigations are no longer valid\footnote{ \label{footnote:uncertainty}
It does not necessarily mean that the asymmetry is not produced for $M<M_{\rm min}$, 
but for the validity of the analysis, we exclude the region from the integral. Since the dominant asymmetry is produced near (or a bit smaller than) $M \sim M_{\rm cr}$, 
the produced asymmetry is not affected by the introduction of the cutoff unless $\mu=T_H$ at the critical mass $M_{\rm cr}$ is smaller than $M_*$.
In the case of $M_{\rm min}< M_{\rm cr}$, our analysis might provide a conservative value of the asymmetry.
}. 
Then changing the integration variable in (\ref{Eq:delta NL}) from $M$ to 
$ X= (M/M_{\rm cr})^2 $, $\delta N_L$ becomes
\al{\label{deltaNL}
\delta N_L&={1\over2}\paren{M_{\rm cr}\over M_P}^2
f(X_{\rm min}, X_0) .
}
The function $f(X_{\rm min}, X_0)$ is defined by an integral 
$
f(X_{\rm min}, X_0) =
\int^{X_0}_{X_{\rm min}} dX
g(X),$
where the integrand is given by 
\al{
g(X)=
{
\dfrac{g_L}{\pi^2}\paren{-\rm{Li}_3 \paren{ -e^{-1/X^{3}} }+\rm{Li}_3 \paren{-e^{1/X^{3}} } }
\over
 \dfrac{7\pi^2g_{\rm other}}{240} + \dfrac{ 3 g_L}{\pi^2}\paren{-\rm{Li}_4 \paren{-e^{-1/X^{3}}}-\rm{Li}_4 \paren{-e^{1/X^{3}} } }
},
\nn
}
and
\al{
X_0 = \paren{ M_{\rm PBH} \over M_{\rm cr} }^2, \ \ \ 
X_{{\rm min}}=
{\rm max}
\paren{
   \paren{ {M_P^2\over M_{\rm cr} M_*}}^2,  \paren{M_P^2\over M_{\rm cr}M_*}^{2\over7}    } .
}
Note that, from Eq.(\ref{criticalmass}), $X_{\rm min}$ is given by
\al{\label{Eq:Xmin n=1}
X_{{\rm min}}&=\begin{cases}
			\paren{\dfrac{M_P^2}{M_{\rm cr} M_*}}^2	\geq  1 &	
				\text{for $M_*<\dfrac{M_P}{256 \pi^3 \sqrt{6\alpha}}\sim10^{-3}M_P$,}\\ \\
			\paren{\dfrac{M_P^2}{M_{\rm cr}M_*}}^{2\over7}  \leq 1 &	
				\text{for $M_*>\dfrac{M_P}{256  \pi^3 \sqrt{6\alpha}}$ \ .}	
		\end{cases}		
}
The function $g(X)$ is numerically depicted  in Fig.~\ref{Fig:integral}.
%
One can see that it is peaked around $X=\mathcal{O}(1)$, which indicates that  the dominant lepton number asymmetry is produced when the BH mass is comparable with the critical mass. 
The damping behavior  of $g(X)$ for large values of $X$ means that when the BH mass is larger than $M_{\rm cr}$, the radiation is almost symmetric  and
never contributes to the lepton number. 
On the other hand, the damping in small values of $X$ implies that the chemical potential of leptons is too large, and the number of emitted leptons is significantly suppressed.
See also App.~\ref{App:analytic} for the behavior and analytical approximations of $f(X_{\rm min},X_0)$ and $g(X)$.

\subsection{Lepton asymmetry in the universe}
In order to discuss the lepton asymmetry in the universe, 
we briefly discuss the cosmological history of the light PBHs.
When the density perturbation becomes as large as  the order one $\delta \rho /\rho \sim 1$, the PBH can be  formed.
The mass of PBH is determined by the energy within the Hubble horizon, namely,
\al{
M_\text{PBH} \simeq  4 \pi \gamma M_P^2H_\text{ini}^{-1} .
}
where $\gamma$ is a numerical factor depending on details of the gravitational collapse. For it is usually considered to be $\gamma  \lesssim 0.2$,
we take $\gamma=0.2$ for simplicity. 
Here $H_\text{ini}$ is the Hubble parameter at the time of the PBH formation.
After it is formed, it emanates the Hawking radiation and 
 when the Hubble parameter becomes the inverse of the PBH lifetime, 
\al{ \label{Hubble-evaporate}
H_\text{eva}\simeq{M_P^4\over M_\text{PBH}^3}3\paren{8\pi}^2 \alpha
,
}
the PBH evaporates completely.　Thus the ratio of the Hubble parameters is given by
\al{
\frac{H_{\rm eva}}{H_{\rm ini} }= 48 \pi \alpha \gamma^{-1} \paren{M_{P} \over M_{\rm PBH}}^2
}
If the universe continues to be in the radiation dominated phase, the Hubble parameter is
related to the scale factor of the universe as $H \propto a^{-2}$.
Then the ratio of the scale factors $a$ during the evaporation is given by
\al{
\paren{a_{\rm eva}\over a_{\rm ini}  } =\sqrt{\frac{\gamma}{48 \pi \alpha }} \paren{M_{\rm PBH} \over M_{ P}}
.
}
Since the energy density of the universe changes as
\al{
\rho(t) = \rho_{\rm rad} (t_i)  \paren{a(t_i) \over a(t)}^4 + \rho_{\rm PBH} (t_i) \paren{a(t_i) \over a(t)}^3,
}
the ratio of the energy density of PBHs  $\rho_{\rm PBH}$ to the total energy density of the universe  $\rho_{\rm rad}$ increases as 
the universe expands. Of course, the evaporation transfers the energy from the PBH to the radiation component and the
actual evolution is more complicated~\cite{Barrow:1991dn}. It is not further discussed in the present paper. 

The temperature of radiation just after the PBH evaporation $T_\text{eva}$ can be estimated from
\al{
H_\text{eva}\simeq\sqrt{\pi^2 g_*\over90}{T_\text{eva}^2\over M_P}.
}
Here $g_*$ is the effective degrees of freedom.
Then we have
\al{\label{Teva}
T_\text{eva}\sim 1.1 \times10^{11}\GeV
\paren{\alpha\over 3.5 \times10^{-3}}^{1/2}
\paren{106.75\over g_*}^{1/4}
\paren{10^5M_P\over M_\text{PBH}}^{3/2}.
}
Notice that the typical value of $T_\text{eva}$ is much higher than the electroweak scale for $M_{\rm PBH} \lesssim10^{11}M_P \sim10^5$g, 
and the lepton asymmetry produced by the PBH evaporation can be converted into the baryon asymmetry by the sphaleron process.

We also note that the Hubble parameter $H_\text{ini}$ must be smaller than the Hubble parameter during inflation
$ H_\text{inf} \sim   10^{14} \sqrt{r/0.1} $ GeV, where $r$ is the tensor to scalar ratio.
Thus we have the lower bound on the PBH 
\al{\label{Eq:lower_bound}
M_\text{PBH} \gtrsim \frac{4\pi \gamma M_P^2}{H_{\rm inf}} \sim 6 \times10^4 \sqrt{0.1\over r} M_P .
}
Only PBHs satisfying the condition can be created in our universe.

Having the above cosmological history in mind, we can estimate the lepton asymmetry  after the PBH evaporation,
\al{\label{Eq:lepton_asymmetry}
{n_L\over s}
&=
{n_\text{PBH}\over s}  \delta N_L
= \Omega_{PBH}
{ \rho_{\rm tot} \over s}
{\delta N_L\over M_\text{PBH}} 
}
where $\Omega_{PBH}=\rho_{\rm PBH}/\rho_{\rm tot}$ is the  ratio of the energy density of 
the PBHs to the total energy density at the epoch of evaporation
 which includes the  radiation from PBHs. 
Assuming domination of the radiation after the evaporation, and using $\rho_{\rm tot} /s = 3T_{\rm eva}/4$, (\ref{deltaNL}) and (\ref{Teva}),
we have 
\al{ \label{abundance-result1}
{n_L\over s}
& \simeq
8.7 \times10^{-9} \ 
\paren{106.75\over g_*}^{1/4}
\paren{\alpha\over 3.5\times10^{-3}}^{5/6} 
\nn
& \times \Omega_\text{PBH} \  f(X_{\rm min},X_0) \
\paren{10^5M_P\over M_\text{PBH}}^{5/2}
\paren{10^{-2}M_P\over M_*}^{4/3}
,
}
which is conserved until now under the assumption that there is no other entropy production. 
From this rough estimation,
we can see that the observed amount of the asymmetry, $n_L/s\sim10^{-10}$, can be successfully produced unless $f(X_{\rm min},X_0)$ is too small.
\begin{figure}
\begin{center}
\hfill
\includegraphics[width=.49\textwidth]{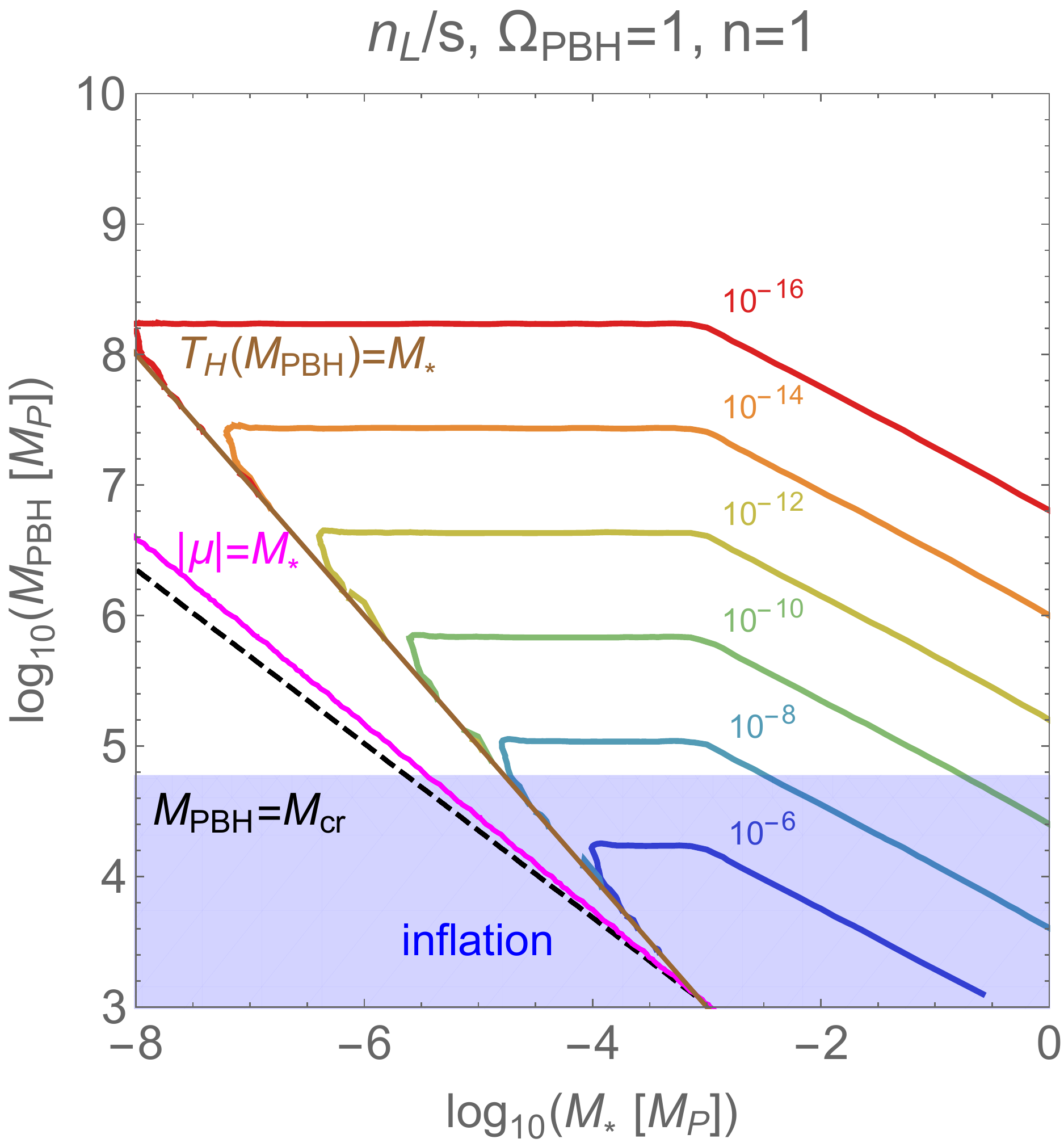}
\hfill
\includegraphics[width=.49\textwidth]{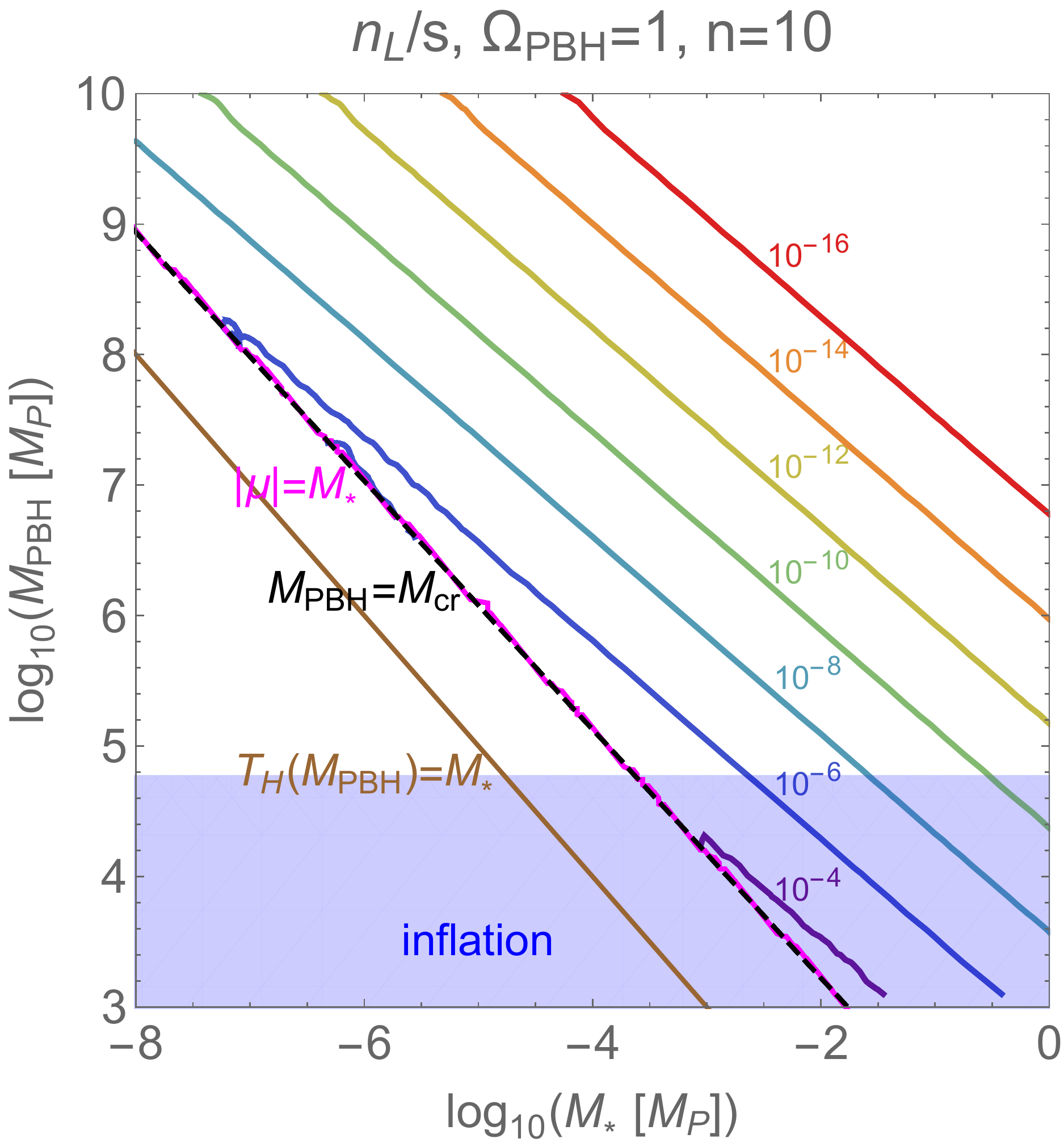}
\hfill\mbox{}
\end{center}
\caption{
The contour plot of the lepton asymmetry produced by the evaporation of the PBH.
The observed baryon asymmetry is $n_B/s\simeq8.7\times10^{-11}$.
The left panel shows the result (\ref{abundance-result1}) for $\Omega_{\rm PBH}=1$.
The dotted line expresses the value of $M_{\rm cr}$ as a function of $M_*$, 
and the brown (and magenta) line represents the condition of $T_H=M_* $ (and $|\mu|=M_*)$, 
below which the typical energy scale becomes higher than the scale $M_*$ and the present calculation is no longer valid. 
The plot is drawn using the formulae  in Appendix \ref{App:analytic}. 
Below $M_*<10^{-3} M_P$, the produced asymmetry becomes flat and independent of $M_*$.
It is due to our prescription to cut off the $X$ integral at $X_{\rm min}$ (see the first paragraph of Appendix \ref{App:analytic} and \eqref{Eq:Xmin n=1}), 
and the asymmetry in this region may be interpreted as a conservative estimation, as noted in footnote~\ref{footnote:uncertainty}.
A PBH with mass lower than  $M_{\rm PBH} \sim 10^5 M_{P}$ is not created in our universe as discussed in 
(\ref{Eq:lower_bound}). 
The graph shows that the baryogenesis  from the PBH works as far as $\Omega_{\rm PBH} > 10^{-2}$. 
The right panel shows the lepton asymmetry  in the case of a generalized CP-violating operator for  $n=10$ discussed in  section \ref{sec:generalCP}.
We use Eq.~\eqref{Eq:lepton_asymmetry2} with  $a_n=1/n$ and $\Omega_\text{PBH}=1$. 
For $n \sim 10$, more asymmetry is efficiently generated compared to the $n=1$ case in the left panel, and the density ratio
of PBH can be as low as $\Omega_{\rm PBH}=10^{-6}$. 
Below the dashed line, the asymmetry is suppressed because the function $g_n(X)$ significantly decreases for $X\lesssim1$.
The magenta line ($|\mu|=M_*$) almost coincides with the dashed line..
Unlike the left panel, the produced asymmetry does not become flat because, within the region of $M_*$ in the graph,
$X_{\rm min}<1$ is always satisfied and the integral is independent of the lower cutoff $X_{\rm min}$.
}
\label{Fig:asymmetry}
\end{figure}

In the left panel of Fig.~\ref{Fig:asymmetry}, we plot the lepton asymmetry generated in presence of the CP-violating interaction 
\eqref{Eq:operator} for $\Omega_{\rm PBH}=1$ 
as a function of $M_*$\footnote{The definition of $M_*$ should be understood as a renormalized one by the effect discussed in the second paragraph of Section \ref{Sec:conclusion}.}. 
In order to generate the observed baryon asymmetry  $n_B/s\simeq8.7\times10^{-11}$, 
the mass of the PBH (the vertical axis) and the scale $M_*$ suppressing the interaction must be on the line with 
$\sim 10^{-10}$.
If the density ratio of PBHs is less than 1, i.e., $\Omega_{\rm PBH}=10^{-s}$ with $s>0$, the parameters $(M_{\rm PBH}, M_*)$ must be on the line with a larger
value $\sim 10^{s-10}$.  
There are 3 lines in the figure.
On the dotted line (the lowest line), the initial mass $M_{\rm PBH}$ of PBH is equal to the critical mass $M_{\rm cr}$. 
The other two lines represent $|\mu|=M_*$ and $T_H=M_*$. 
The region below these lines is beyond the reach of this paper 
since the typical energy scale is larger than $M_*$. 

From Eq.(\ref{Eq:Xmin n=1}), we see that the value
$M_*\sim10^{-3}M_P$  corresponds to $X_{\rm min}\sim1$, and the $M_*$ dependence of asymmetry becomes
different between the right region  with $X_{\rm min}\lesssim1$ and the left region with $X_{\rm min}\gtrsim1$.
For $X_{\rm min}\lesssim1$,
according to Fig.~\ref{Fig:integral}, $f$ (an integral of $g$) becomes almost independent of $X_{\rm min}$.
Since the asymmetry is most dominantly generated around the dotted line ($X=1$) and 
the critical mass increases as $M_*$ decreases (see Eq.(\ref{criticalmass})),
 the asymmetry also increases when $M_{\rm PBH}$ fixed. 
On the other hand,  for $X_{\rm min}\gtrsim1$, 
$f$ strongly depends on the lower cutoff $X_{\rm min}$.
It is shown in the first paragraph of Appendix \ref{App:analytic} that the produced asymmetry becomes independent of $M_*$ for $X_{\rm min}\gtrsim1$.
This is the reason why the generated asymmetry becomes flat as a function of $M_*$ below $M_* \sim 10^{-3} M_{\rm P}$. 

The shaded region is not allowed because below $M_{\rm PBH}\sim 10^5$ g, PBH is not created in our universe (\ref{Eq:lower_bound}).
Therefore,  Figure \ref{Fig:asymmetry}  shows that the baryogenesis  from the PBH works as far as $\Omega_{\rm PBH} > 10^{-2}$ 
for the simplest CP violating operator of dimension 8. It is based on our  conservative assumption\footnote{The assumption is partially based on the fact that
the typical energy of the Hawking radiation is given by $T_H (>|\mu|)$ and thus number of emanated particles is drastically reduced for very high $T_H$.}
 that the lepton asymmetry produced
in the region of $T_H>M_*$ is not counted.

\subsection{More general CP-violating interactions}  \label{sec:generalCP}
So far we have studied the CP-violating interaction of \eqref{Eq:operator}.
We extend it to  more general higher dimensional operators introduced in Eq.~\eqref{Eq:operator2}\footnote{A truncation of higher order terms is assumed  in this subsection. 
In order to estimate the asymmetry starting from the ultraviolet theory such as the model in Sec.~\ref{Sec:higher_dimensional_operators}, it is necessary to compute the full propagator in curved background without relying on the derivative expansion.
 We want to come back to this problem in future publications.
}.
The calculation is the same as the simplest case discussed so far.
The chemical potential  is dynamically generated at the horizon and the ratio to the Hawking temperature is given by
\al{ \label{ratio-muTn}
{\mu\over T_H}\bigg|_{r=r_H}
= - \paren{M_{\rm cr}\over M}^{4n+2}
}
where the critical mass is given by 
\al{ \label{ninf-criticalmass}
& M_{\rm cr} =: \sqrt{C_n} M_P, \nn
&
C_n = (n a_n)^{1 / (2n+1)} (128 \alpha \pi^2)^{1/(2n+1)} \paren{ 32 \sqrt{3}  \pi^2}^{2n/(2n+1)} 
\paren{M_P \over M_*}^{4n/(2n+1)} 
.
}
Notice that, compared to the $n=1$ case in (\ref{criticalmass}), the exponent of $(M_P/M_*)$ in $\sqrt{C_n}$ is larger and accordingly
the critical mass becomes larger. It is good for producing larger asymmetry since the asymmetry is produced when the BH mass is around the critical mass.
Parametrizing $M_*$ as $M_*=10^x M_P$ and  $M_{\rm PBH}=10^y M_P$, 
the condition of $M_{\rm PBH} = M_{\rm cr}$ for $n \rightarrow \infty$ with $a_n=1/n$ becomes
\al{
y = 1.37 -x
.
}
On the other hand, as shown in eq. (\ref{ratio-muTn}), the chemical potential becomes too large when $M$ becomes larger than $M_{\rm cr}$. It is bad for  asymmetry generation
because it enhances the typical energy $\langle E \rangle \sim \mu$ of emitted particles and consequently reduces  number of particles $dM/\langle E \rangle$ while the BH decreases its mass by $dM$.  These two effects
compete as $n$ becomes large.

By introducing a new integration variable $X=(M/M_{\rm cr})^2$ as before, 
we obtain the lepton number emitted from a single PBH as
\al{  \label{deltaNLn} 
\delta N_L
& =
{1\over2}\paren{M_{\rm cr}\over M_P}^2 f_n(X_{\rm min},X_0),
}   
where $f_n$ is an integral 
$f_n(X_{\rm min},X_0) =\int^{X_0}_{X_{\rm min}} dX g_n(X)$  over the PBH mass $X$.
Here the function $g_n$ is given by
\al{ \label{gnX}
g_n(X)
=
{
\dfrac{g_L}{\pi^2}\paren{-\rm{Li}_3 \paren{ -e^{-1/X^{2n+1}} }+\rm{Li}_3 \paren{-e^{1/X^{2n+1}} } }
\over
\dfrac{7\pi^2}{240}g_{\rm other}+\dfrac{ 3 g_L}{\pi^2}\paren{-\rm{Li}_4 \paren{-e^{-1/X^{2n+1}}}-\rm{Li}_4 \paren{-e^{1/X^{2n+1}} } }
}
}
and the lower bound of the integral is given by
\al{
X_{{\rm min}}=
{\rm max}
\paren{ \paren{M_P^2\over M_{\rm cr}M_*}^2, \paren{M_P^2\over M_{\rm cr}M_*}^{2\over4n+3}
}.
}
By using Eq.~\eqref{ninf-criticalmass}, it is found
\al{
X_{{\rm min}}&=
		\begin{cases}
			\paren{\dfrac{M_P^2}{M_{\rm cr} M_*}}^2 \geq 1	&	
				\text{for $M_*< M_{*{\rm eq},n}$}:= \paren{n a_n\, 128\alpha\pi^2}^{-1/2}\paren{32\sqrt{3}\pi^2}^{-n}M_P , \\ \\
			\paren{\dfrac{M_P^2}{M_{\rm cr}M_*}}^{2\over4n+3}	\leq 1 &	
				\text{for $M_*>M_{*{\rm eq},n}$}	
		\end{cases}	
}
and, for  $n=10$,  $M_{*{\rm eq},n=10}\sim10^{-28}M_P$.
Thus, in the region of $M_* > 10^{-8} M_P$, $X_{\rm min}<1$ is 
always satisfied.
The function $f_n(X_{\rm min},X_0)$ behaves similarly to the previously defined function $f(X_{\rm min},X_0)=f_1(X_{\rm min},X_0)$, 
but, as shown in Fig.\ref{Fig:integrand approximation}, the function $g_n$ (the integrand of $f_n$) is more sharply peaked  around $X_0 \sim 1.$ 
In the large $n$ limit, 
\al{
\delta N_L
& ={1\over2} (na_n)^{1/2n} 
 \paren{ 32 \sqrt{3}  \pi^2} \paren{M_P \over M_*}^2 f_n(X_{\rm min},X_0) .
}

\begin{figure}
\begin{center}
\hfill
\includegraphics[width=.45\textwidth]{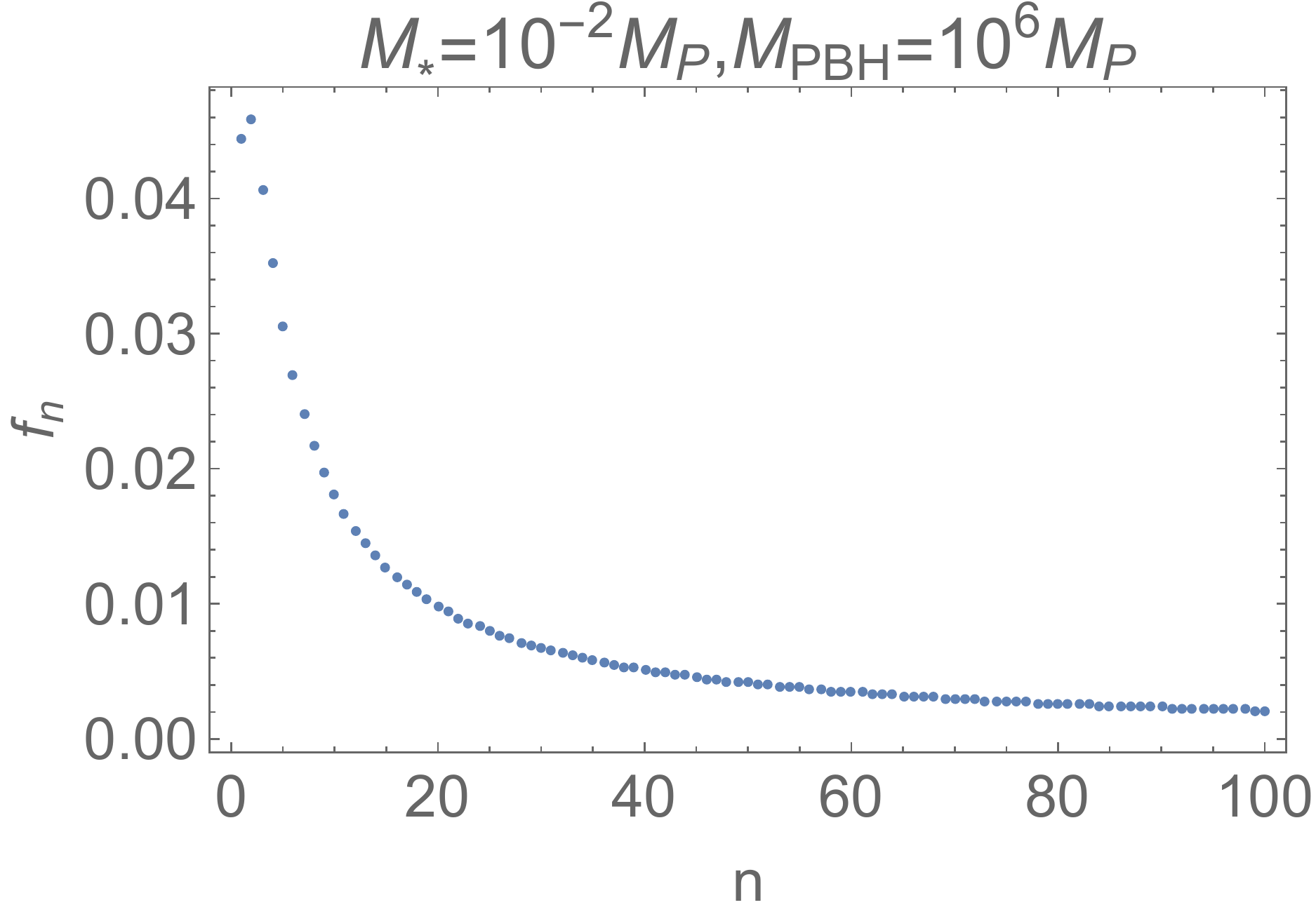}
\includegraphics[width=.49\textwidth]{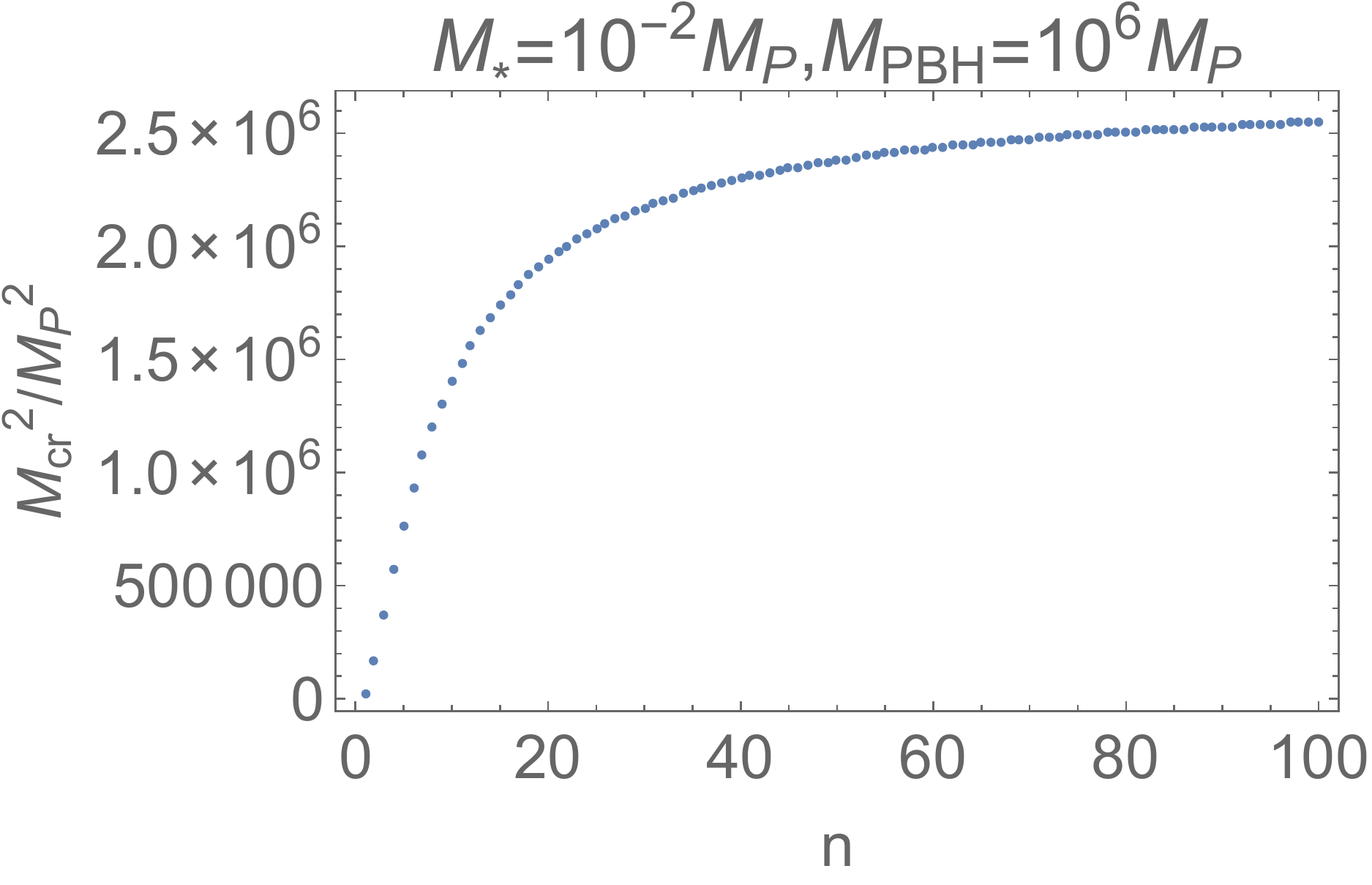}
\hfill\mbox{}
\hfill
\\
\includegraphics[width=.49\textwidth]{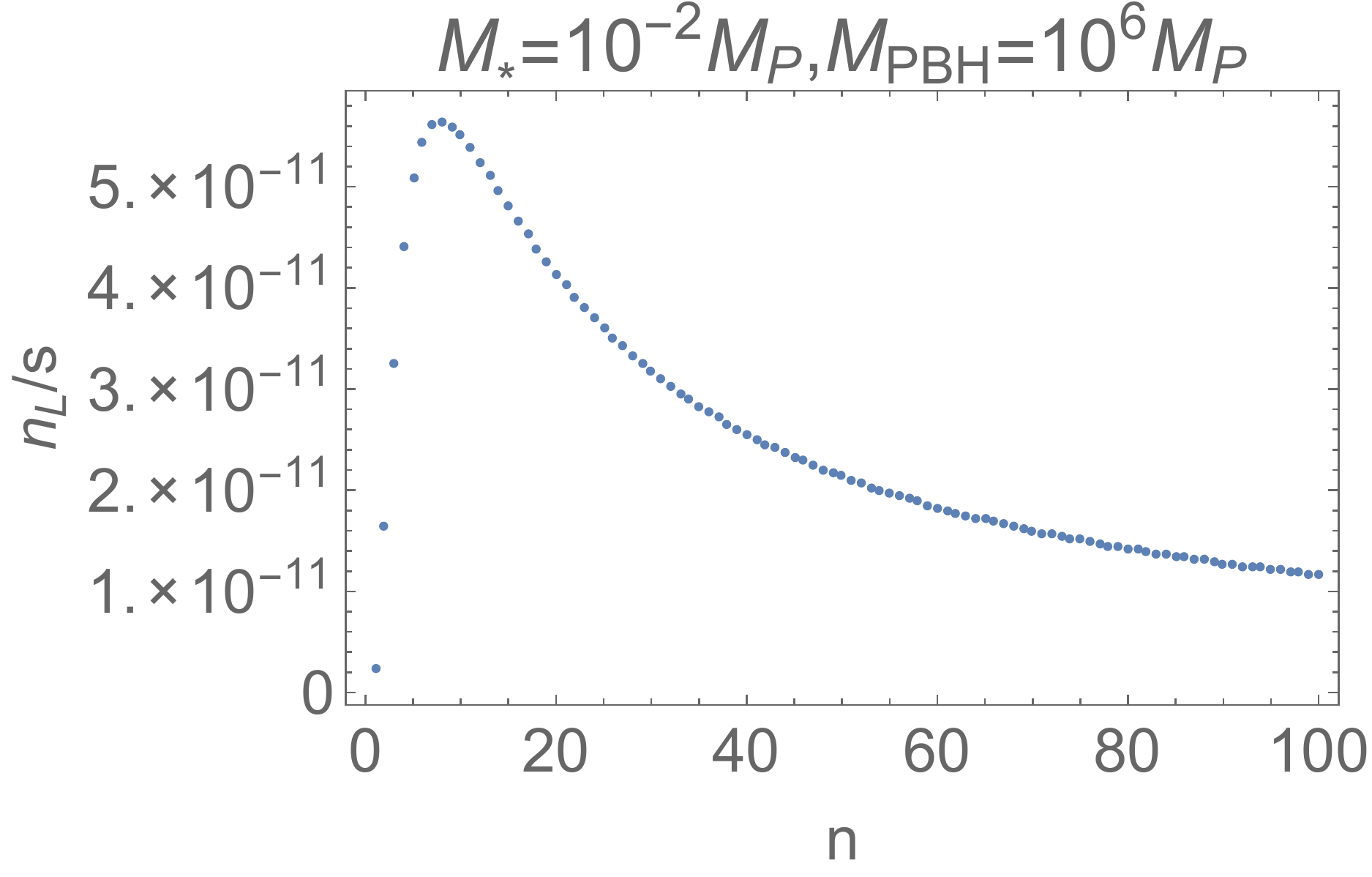}
\end{center}
\caption{
$f_n$, $M_{\rm cr}^2/M_P^2$ and $n_L/s=f_n\times M_{\rm cr}^2/(2M_P^2)$ are plotted as a function of $n$.
Here we take $M_*=10^{-2}M_P, M_\text{PBH}=10^6M_P$.
Although $f_n$ is a decreasing function of $n$ for $n\geq2$, $M_{\rm cr}$ is monotonically increasing and thus $n_L/s$ has a peak  around $n=10\text{--}20$.
}
\label{Fig:n}
\end{figure}

Hence, for large $n$, the lepton asymmetry is found to be
\al{  \label{Eq:lepton_asymmetry2}
{n_L\over s}
& \simeq
9.3\times 10^{-7} \ 
\paren{106.75\over g_*}^{1/4}
\paren{\alpha\over 3.5\times10^{-3}}^{1/2} 
\nn
& \times \Omega_\text{PBH} \  f_n(X_{\rm min},X_0) \
\paren{10^5M_P\over M_\text{PBH}}^{5/2}
\paren{10^{-2}M_P\over M_*}^{2} (na_n)^{1/2n}
.
}
According to the behavior of $a_n$, 
the asymptotic behavior of $(n a_n)^{1/2n}$ is given as
\al{
(n a_n)^{1\over 2n}
\to 
\left\{
\begin{array}{ll}
1  & \text{for  $ a_n \sim \dfrac{1}{n}  $}
\\ \\
\sqrt{e \over n} & \text{for  $ a_n \sim \dfrac{1}{n!}  \sim n^{-n} e^{n}   $}
\end{array}
\right.
.
}
As discussed in App.~\ref{App:analytic}, the function $f_n$ is suppressed by $n$ as $f_n \sim1/n$, and the asymmetry vanishes in the $n \to \infty$ limit.
It is due to the rapid increase of the chemical potential $\mu$ for a larger $n$ at $M<M_{\rm cr}$. 
However, it is interesting to note that, for moderate values of $n$, the higher dimensional operators are more efficient to produce larger amount of the asymmetry thanks to the increasing of $M_{\rm cr}$.
In order to see the competition of these two effects of taking large $n$,
we plot  $n$ dependences of the function $f_n$, the critical mass $M_{\rm cr}^2/M_P^2$ and the Lepton asymmetry  $n_L/s=f_n\times M_{\rm cr}^2/(2M_P^2)$ in Fig.~\ref{Fig:n}.
It can be seen that $f_n$ behaves as $1/n$ for large $n$, as expected.
On the contrary, $M_{\rm cr}$ is a monotonically increasing function of $n$. 
As a result, $n_L/s$ has the peak around  $n=10\text{--}20$.

We now explicitly estimate the produced asymmetry created by the higher dimensional interactions, assuming  $(n a_n)^{1/2n} \sim 1.$
In the right panel of Fig.~\ref{Fig:asymmetry}, we plot the produced asymmetry $n_L /s$ in Eq.~\eqref{Eq:lepton_asymmetry2}
 for $\Omega_\text{PBH}=1$ and $n=10$. 
 We can see that the observed asymmetry, $n_B/s\simeq8.7\times10^{-11}$, is realized in a wide range of parameters 
as far as $\Omega_{\rm PBH} > 10^{-6}$.

We note that, if the mass of the PBH is less than $10^{8}$g~\cite{Carr:2009jm}, 
 there are no observational  constraints.  Thus the mass region in which the present baryogenesis scenario works is all allowed. 
  It is also interesting to note 
 that the generalized higher dimensional interactions suppressed by  a power of $M_*$ can generate larger baryon asymmetry.
 It is due to the fact that the critical mass $M_{\rm cr}$, around which the most lepton number is generated,  becomes larger for larger $n.$   
But it  also indicates that the present calculation of using the derivative expansions needs improvement.
 Actually, when the BH mass is reduced to the critical mass $M_{\rm cr}$, the scale of the curvature tensor becomes comparable
 to $M_*$, and the Compton wave length of  the underlying particles that induce the CP violating coupling to the background gravity becomes
 comparable to the horizon radius of the BH. 
 It is  important  to study the asymmetric Hawking radiation  without resorting to the derivative expansions adopted in the present paper.
 We hope to come back in future investigations.

 \section{Possible origin of CP-violating operators}\label{Sec:higher_dimensional_operators}
We discuss a possible origin of the CP-violating  higher dimensional  operators through interactions with 
the right-handed neutrinos. 
Note that we have neither succeeded to obtain such interactions by explicit calculations or
the mechanism discussed in the previous section is  not restricted to this particular possibility.
But, in order to get a rough order estimation of the scale of $M_*$, 
we study this possibility in this section\footnote{If the right-handed neutrinos are responsible for the CP-violating higher dimensional interactions, 
the ordinary leptogenesis can also occur and we need to take it into account to calculate the total abundance of baryon asymmetry.
In the present paper, we consider only the abundance generated through the PBHs. }. 

\begin{figure}
\begin{center}
\hfill
\includegraphics[width=.6\textwidth]{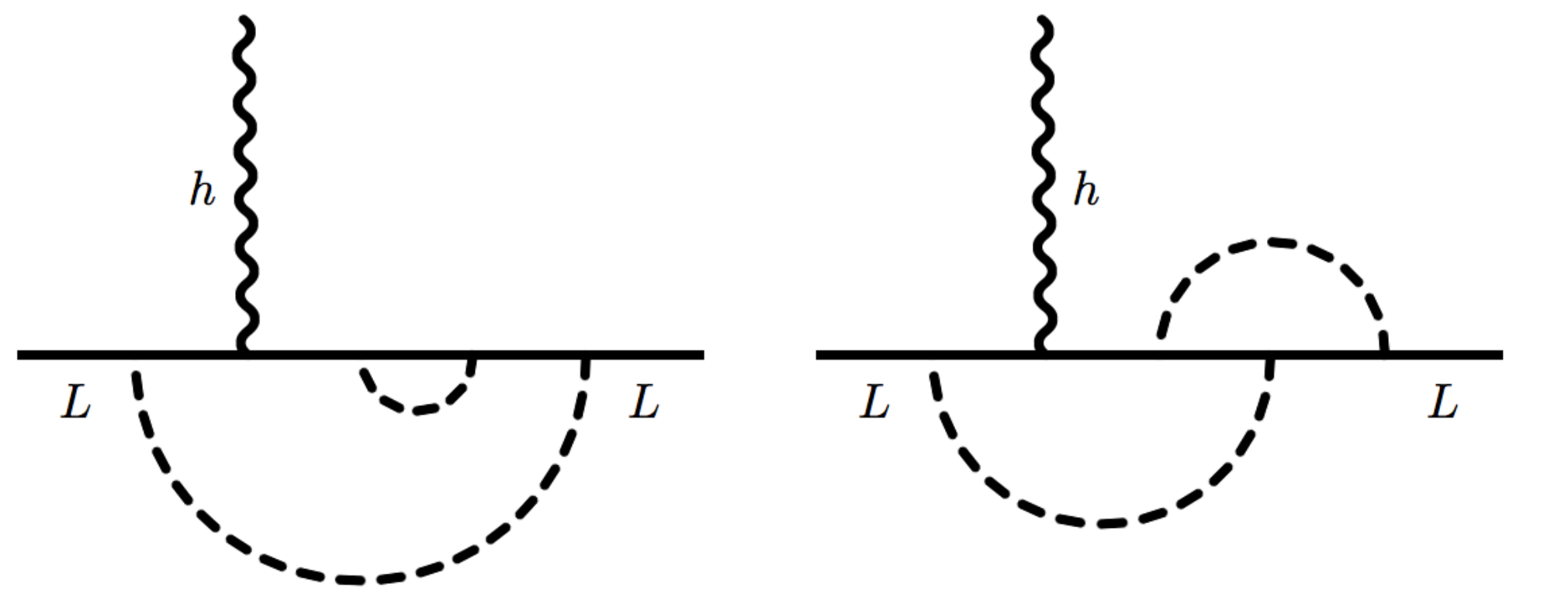}
\hfill\mbox{}
\end{center}
\caption{
The diagrams inducing the effective operator Eq.~\eqref{Eq:Mcdonald_Shore} by integrating out the right handed neutrinos.
$L$ and $h$ are the lepton doublet, and external gravitational field, respectively.
}
\label{Fig:diagrams}
\end{figure}

In order to generate these operators, 
the underlying physics need to  violate CP symmetry and the lepton number.
One of the simplest possibility is  the Majorana mass terms of the
right handed neutrinos. As usual,  imaginary phases of the mass matrix break the  CP symmetry.
The Lagrangian we consider is 
\al{\label{Eq:seesaw}
\mathcal{L}
=
\mathcal{L}_\text{SM}
+y_N\overline{L}{N_R}\widetilde{\Phi}
+\frac{M_{R}^{}}{2}\overline{{N^c_R}}{N_R}_i+\text{h.c.},
}
where $y_N$ is the neutrino Yukawa coupling, and $M_N$ is the mass of the  right handed neutrinos.
Recently, Mcdonald and Shore~\cite{McDonald:2015ooa,McDonald:2015iwt}
explicitly  calculated loop diagrams containing the above interactions coupled with the gravity in the external lines.
They took the conformally flat metric,  
$g_{\mu\nu}=(1+h)\eta_{\mu\nu}$ where the field $h$ is treated as a background field, and then calculated
the two-loop Feynman diagrams in Fig.~\ref{Fig:diagrams} to obtain 
the higher dimensional operator used in the  gravitational leptogenesis,
\al{\label{Eq:Mcdonald_Shore}
{1\over M_{MS}^2}\partial_\mu \mathcal{R} j^\mu
.
}
Since the right handed neutrinos are integrated out, 
the mass scale of the higher dimensional interaction is essentially given by\footnote{These authors further claimed possible appearance of 
an enhancement factor depending on the mass hierarchy of the right handed neutrinos.  We do not discuss such effects here.}
\al{
{1\over M_{MS}^2}
\sim
{1\over(16\pi^2)^2}{\text{Im}((y_{N}y_{N}^\dagger)^2)\over M_N^2}.
}

%
The higher dimensional operators utilized  in the present paper will also arise by calculating similar diagrams with more gravitons 
in the external lines \footnote{It is not evident what types of higher
dimensional operators can be induced, but their specific forms are not very important in our discussions as commented in the footnote \ref{footnote:Gauss}. }.
To prove it,  we need to consider a more generic form of the metric and further involved calculations are necessary, so we leave it for future investigations.
From dimensional and diagramatic arguments, it can be expected that the mass scale $M_*$ is given by
\al{
{1\over M_*^4}
\sim
{1\over(16\pi^2)^2}{\text{Im}((y_{N}y_{N}^\dagger)^2)\over M_N^4}.
}
Let us now give a rough estimation for the scale $M_*$ using this formula.
The see-saw mechanism indicates that the typical order of $M_N$ is given by $M_N  \sim (y_N v)^2/m_\nu \sim y_N^2 10^{16}$ GeV,
where $v \sim 100$ GeV and $m_\nu \sim 10^{-3}$ eV. The scale $M_*$ is proportional to $M_N$ but also dependent on the Yukawa coupling $y_N$
and the CP-violating parameter $\epsilon:=\text{Im}((y_{N}y_{N}^\dagger)^2/|(y_{N}y_{N}^\dagger)^2|)$. If we take the Yukawa coupling as $y_N \sim 1$, smallness of $\epsilon$ makes the scale $M_*$ 
larger than $10^{16}$ GeV. On the other hand, if we assume that $y_N$ is as small as, e.g. $y_N =10^{-5}$ and the CP-violating parameter is of order 1,
the scale $M_*$ can be lowered to $10^{11}$  GeV. 
Note that from recent arguments on the hierarchy problem of the electroweak scale, it is preferable to take the scale of the right-handed neutrinos relatively lower. 
Hence the scale  $M_*$ may be expected to take a value in the region, $10^{-7} M_P < M_* < M_P.$ 

We also emphasize that, other than the right handed neutrinos, 
if CP is broken in the UV theory including gravity, 
we can naturally expect  appearance of the operators like Eq.\eqref{Eq:operator2}.
These  are left for future studies.

\section{Effect of washout outside the BH} \label{Sec:washout}
If the CP-violating interaction such as Eq.\eqref{Eq:operator2} is present, decay of the PBH mass by Hawking radiation 
breaks time-reversal symmetry and generates chemical potential at the horizon. 
Then the radiation becomes asymmetric. As already mentioned before, the distribution looks thermal but it does not mean that 
the lepton number violating interaction is in the thermal equilibrium. 
The Hawking radiation is emitted because of the quantum mechanical effect, and reflects the fact that
the quantum vacuum near the horizon is different from the Minkowski vacuum.
In this respect, the mechanism of baryogenesis is very different from the gravitational baryogenesis \cite{Davoudiasl:2004gf}, 
in which decoupling of the baryon number violating interaction is necessary to fix the final amount of asymmetry.

It is, however, necessary to check whether the generated asymmetry is washed out by the baryon (or lepton) number violating
interactions outside of the black hole horizon. 
As discussed in the previous section, we assumed that the microscopic process violates the lepton number 
through the interaction with the right-handed neutrinos. 
By integrating out them, we have the  following dimension-5 operator,
\al{
S_{\slashed L}&=
\int d^4x \paren{{1\over\Lambda}LLHH} .
}
%
After the Higgs acquires the vev $\langle H \rangle=v \simeq246\GeV$, it gives the mass to the neutrinos.  Hence $\Lambda$ is determined to be
\al{
\Lambda \sim {v^2\over m_\nu} \sim 2.5 \times 10^{-2} M_P,
}
where $m_\nu$ is the neutrino mass and we put $m_\nu=10^{-3}$ eV. 

The cross section is estimated by $\sigma \sim 1/\Lambda^2$ at low energy scale. The particle number density is given by $n \sim T_H^3$ near the horizon
where the particles are created. 
Then the scattering rate of the lepton number violating interaction at the horizon is given by 
\al{
\Gamma_{\slashed L}^0 = \sigma n =T_H^3/ \Lambda^2  \sim 1.6 \times 10^{-12} \paren{10^5 M_P \over M}^3 M_P.
}
But the rate is much slower than  the typical  time scale of the created particle to move away from the BH, 
namely $\Gamma_{\slashed L}^0  \ll r_H^{-1} \sim (M_P/M) M_P$
for the BH mass $M \gtrsim 10^{5} M_P$. 
Hence the particle density is quickly diluted by the factor $(r_H/r)^2$, where $r$ is the distance from the BH, and 
the interaction rate is reduced to $\Gamma_{\slashed L}=\Gamma_{\slashed L}^0 (r_H/r)^2.$
Since the Hubble parameter of the universe at the epoch of evaporation in (\ref{Hubble-evaporate}) is estimated to be
$H_{\rm eva} \sim 10^{-14} (10^5 M_P/M)^3 M_P$,
the dilution factor $(r_H/r)^2$ instantly makes $\Gamma_{\slashed L}$ much smaller than $H_{\rm eva}$. 
Hence the interaction is not in the chemical equilibrium at the time of evaporation with the Hubble $H_{\rm eva}$.

Next we check the condition of washout 
when the particles are at higher energy than $M_N$. 
The infinite blue shift near the BH horizon enhances  the energy of the particle by the factor $(1-r_H/r)^{-1/2}$.
Then, near the horizon,  the cross section of lepton number violating interaction is replaced by its high energy counterpart
 $\sigma \sim y_N^2/s \sim y_N^2 (1-r_H/r) /T_H^2.$
Taking the dilution factor $(r_H/r)^2$ into account, we have
  \al{
 \Gamma_ {\slashed L}  \sim  y_N^2 T_H {r_H^2\over r^2} (1-{r_H \over r}) .
  }
The maximal value of $\Gamma_{\slashed L}$ is obtained  by
\al{ \label{maxGamma}
 \Gamma_{\rm max} \sim 0.15 \  y_N^2 T_H = 0.15 { y_N^2  \over 4 \pi r_H}
}
at the position $r \sim 1.5 \ r_H$.
Again this is much smaller than $r_H^{-1}$ by a factor $10^{-2} y_N^2$, and 
the emitted particles move away quickly from the BH, so the scattering rate is reduced to 
$\Gamma_{\slashed L} \sim (r_H/r)^2 \Gamma_{\rm max}$. Compared with the Hubble $H_{\rm eva}$ in  (\ref{Hubble-evaporate}),
when the particle moves to $r \sim 10^3 r_H$, $\Gamma_{\slashed L} < H_{\rm eva}$ is satisfied. 
Hence, if $y_N <0.1$ the time that the particle moves to that position  is sufficiently  short for the lepton number violating 
interaction to occur, and  the generated lepton number is never washed out. 

%
\section{Summary and discussions}   \label{Sec:conclusion}
In this paper, we have proposed a new scenario of baryogenesis from evaporating PBHs. 
The key element is the CP violating operator Eq.~\eqref{Eq:operator}, which generates the splitting of energy spectrum between particles and anti-particles
if the BH is decaying.
The mechanism is similar in spirit to the gravitational baryogenesis \cite{Davoudiasl:2004gf}  or the mechanism by Hook  \cite{Hook:2014mla}  
who applied the gravitational baryogenesis to the PBH, but an essential difference is that we make use of the time-evolution of the BH itself \cite{Banks:2015xsa},
not the cosmological evolution, to generate the chemical potential. 
Because of this, the ratio of the chemical potential to the Hawking temperature $\mu/T_H$ becomes a function of the decaying mass of evaporating BHs
and, when the mass becomes under the critical mass, the ratio $|\mu/T_H|$ exceeds 1. After this epoch, maximal asymmetry can be generated.
Due to such efficiency of generation mechanism, even though the CP-violating operator is largely suppressed by high energy scale $M_*$,
we show that  sufficient amount of baryon asymmetry $n_B/s\simeq8.7\times10^{-11}$
can be obtained in wide range of parameter space if the density ratio of PBHs at the epoch of evaporation is $\Omega_{\rm PBH} > 10^{-6}$.
If the scenario really explains the baryon asymmetry of the universe, it constrains model buildings beyond the standard model
because the PBH can radiate heavy particles that may decay later, e.g. during the BBN. Thus the present scenario
favors simpler model buildings such as \cite{Asaka:2005an}.
 
The estimation of the generated asymmetry and the requirement for $\Omega_{\rm PBH}$ in the present paper will change if we take
other effects into account. In our analysis we evaluated the chemical potential at the horizon $r=r_{H}$. There is, however, a discussion
\cite{Unruh:1977ga} \cite{Giddings:2015uzr}
that the Hawking radiation originates in the larger-region $r =c r_H> r_H$. 
According to \cite{Giddings:2015uzr}, $c=3\sqrt{3}/2 \sim 2.6$ is given for high frequency modes.  
But the effect can be always absorbed in the definition of the scale $M_*$.
Indeed if we instead evaluate $\mu$ at $r=c r_H$, 
 the chemical potential is reduced by $c^{-6n}$ and the critical mass $M_{\rm cr}$ is reduced by the factor $c^{-6n/(4n+2)}$.
From the definition of the critical mass, it is equivalent to increasing  the effective $M_*$ by $c^{6n/(4n+2)}$. 
For $n=1$ and $n=10$ with $c=3\sqrt{3}/2 $, the numerical factors are $2.6$ and $3.9$ respectively.

In the analysis we used the adiabatic approximation, namely we have implicitly assumed that 
the time scale  characterizing the Hawking radiation is shorter than that of the change of mass of the PBH.
Here we confirm the validity of this assumption.
The typical time scale of Hawking radiation is estimated from the uncertainty relation between  time and energy,
\al{
\Delta t
\sim
{1\over  \Delta E}
\sim
{1\over T_H}.
 }
 %
The assumption of the adiabaticity is justified if the change of the PBH mass  is negligibly small during $\Delta t$.
Using Eq.~\eqref{Eq:mass_change}, we have
\al{
\left|
{dM\over dt} \Delta t
\right|
&\simeq
{\pi N\over480}
{M_P^4\over M^2 T_H}
\nn
&
=
{\pi N\over480}
\paren{{M_P\over M}}^2M
}
which is much smaller than $M$ for $M\gg M_P$.
Therefore, we can safely treat the system adiabatically.
 
\begin{figure}
\begin{center}
\hfill
\includegraphics[width=.6\textwidth]{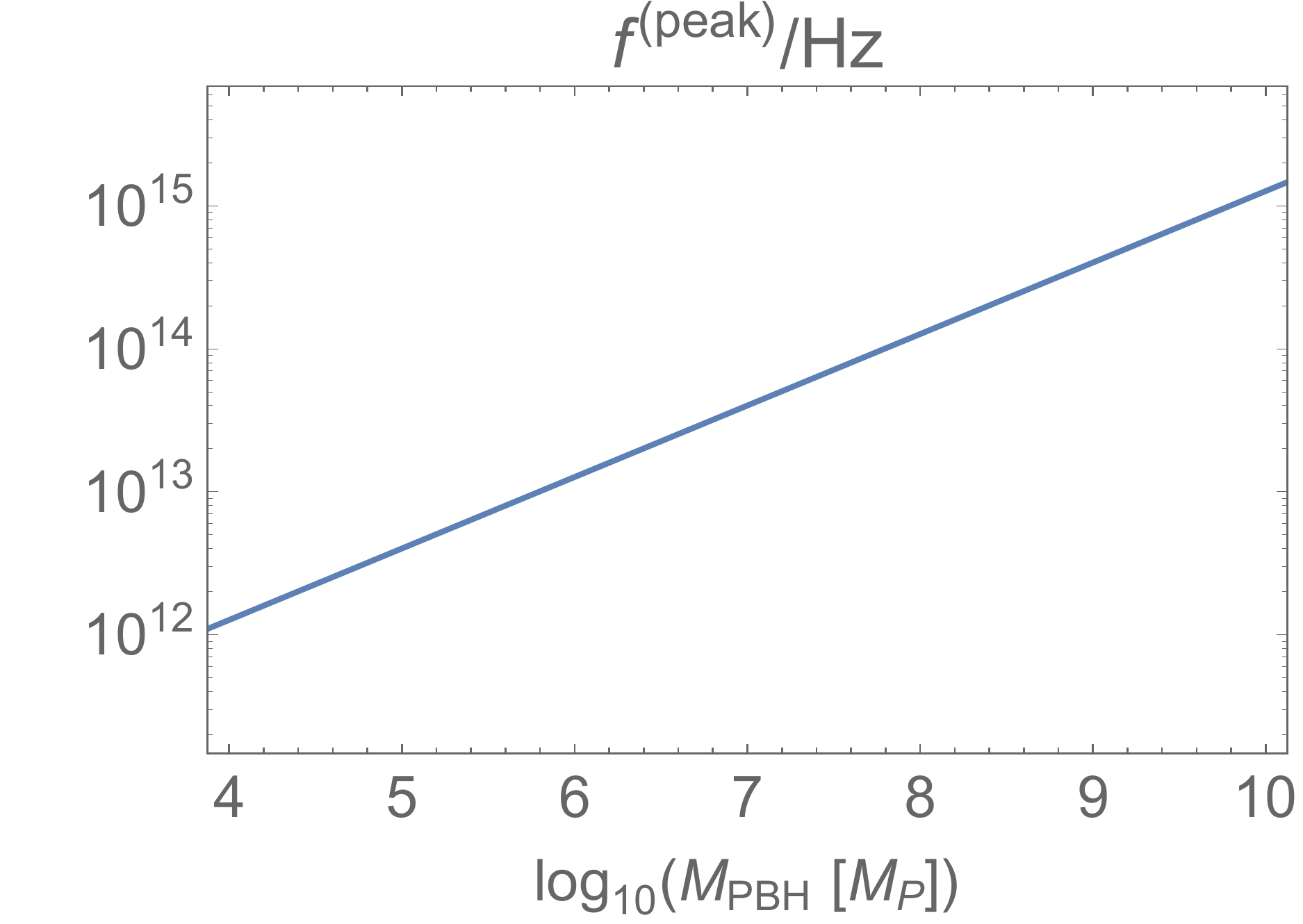}
\hfill\mbox{}
\end{center}
\caption{
The peak frequency of gravitation wave from Hawking radiation from PBH as a function of $M_\text{PBH}$. 
}
\label{Fig:fpeak}
\end{figure}

If PBHs are responsible for the baryon asymmetry in the universe, they can also be responsible for gravitational waves.
PBHs emit  gravitational waves either by Hawking radiation or by a formation of PBH binaries, but
the Hawking radiation would provide the strongest signal~\cite{Dolgov:2011cq}.
In Ref.~\cite{Dolgov:2011cq}, they estimated the peak frequency $f^\text{(peak)}$ and the peak amplitude $h_0^2\Omega_\text{GW}$ as
\al{&
f^\text{(peak)}
=
4\times10^{12}\,\rm{Hz}\,
\sqrt{
M_\text{PBH}\over 10^5 M_P},
&
h_0^2\Omega_\text{GW}
\sim
10^{-7},
}
if PBH dominates the universe.
Note that the peak amplitude does not depend on $M_\text{PBH}$. 
In Fig. \ref{Fig:fpeak}, the peak frequency of the gravitational waves from the PBH is plotted as a function of its mass.
It is amusing to consider a cosmological history  in which PBHs dominate in the early universe. But
the expected gravitational waves seem to be difficult to observe since its frequencies are too high for the
near future experiments. 

The final comment is the formation mechanism of the PBHs. There are various possibilities to create PBHs in the early universe 
as summarized in the introduction.
The bubble collisions~\cite{Crawford:1982yz} associated with the  first order phase transition in the universe~\cite{Hawking:1982ga}
is becoming more interesting recently since the  discovery of the Higgs boson and strong constraints on the TeV scale physics 
have stimulated  reconsideration of our view of the cosmological history. In particular,  revival of radiative symmetry breaking via the Coleman
Weinberg mechanism ~\cite{Iso:2009ss,Iso:2009nw,Iso:2012jn} suggests 
that the  phase transition will be a strong first order type. 
In such models,  bubble collisions of the true vacua can generate strong gravitational waves \cite{Jinno:2016knw}, topological objects such as monopoles
\cite{Khoze:2014woa}, or PBHs ~\cite{Crawford:1982yz}. 
It is interesting to pursue further cosmological consequences of the strong first order phase transitions, in relation to the 
particles physics models beyond the SM.

\appendix 
\section{Is chemical potential physical?}  \label{App:chemical potential}
If  the (lepton number) current is conserved, 
the operator of the type in Eq.~\eqref{mostgeneralop} seems to vanish by integrating by parts or by rotating the phase of leptons
as  $\psi$ as $\psi\to\exp(i F(\mathcal{R_{....}}))\psi$. 
It implies that the lepton number violation is necessary to justify the microscopic origin of such CP-violating effective operators.
Indeed McDonald and Shore \cite{McDonald:2015iwt} obtained such a term by integrating the right-handed neutrinos whose interactions violate the CP and 
the lepton number conservation  microscopically.

But it is still paradoxical why we are not allowed to do such phase rotation to remove the effect of lepton asymmetry. 
First note that if the interaction is replace by $\mu J^0$ where $\mu =\dot{F}$ is assumed to be a constant,
the phase rotation becomes $\psi \to \exp (i \mu t) \psi$.  This is nothing but the shift of the energy $\omega \to \omega -\mu$
 for leptons and $\omega \to \omega +\mu$ anti-leptons. Since the vacuum state is defined 
 so as to fill  all negative energy levels, the phase rotations simply changes the definitions of the vacuum. 
 Thus such phase rotation should be correlated with the definition of the vacuum state, and with the 
 definition of the lepton number of the vacuum.
 
 In the case we are discussing, the current is coupled to the total derivative $\partial_\alpha F(\mathcal{R})$ and $F(\mathcal{R})$
 is a smooth function in the spacetime. Especially it vanishes at $r=\infty$ and takes a nonvanishing value at the BH horizon $r=r_H.$
 Therefore, if we assume that $F(\mathcal{R})$ is a slowly changing function in time and can be expanded as $F(t,r)=F_0(r)+ \mu(r) t + \cdots$, 
  the phase rotation changes the energy levels of leptons $\omega \to \omega -\mu(r)$ as a function of the position $r.$
 In this sense, the situation is similar to the discussion of chiral anomaly as the spectral flow in the Hamiltonian formulation.
 Now the question is that how we can define the vacuum of quantum field at $r=\infty$ and $r=r_H$ separately. 
 And this is nothing but the issue of the Hawking radiation. 
 At $r=\infty$, $F(\mathcal{R})$ vanishes and there is no ambiguity in defining the vacuum. 
 At $r=r_H$, we first define the appropriate vacuum state (such as the Unruh vacuum) so that an infalling observer does not encounter
 any divergences,  and then calculate the effective action.
 Once the effective operator is induced as performed in \cite{McDonald:2015iwt, McDonald:2015ooa}, 
 then we can no longer shift the energy level (see also \cite{Dolgov:1980gk}  and \cite{Toussaint:1978br}.) 
 This is the reason why we should not rotate the basis after we calculate the effective interaction, and 
 the CP-violating operator and the  resulting chemical potential derived from Eq.~\eqref{Eq:seesaw} 
  has a physical meaning.\footnote{From a different point of view, Eq.~\eqref{Eq:operator} would be meaningful since this operator contains running coupling $y_N$. 
Because this coupling depends on scale, the operator cannot be absorbed by fermion rotation in a covariant way.}
It will be interesting to obtain the asymmetry in spectrum of 
the Hawking radiation without resorting to the calculation of the effective interaction, i.e. by explicitly 
calculating the lepton wave function in the eikonal approximation with the coupling to the right-handed neutrinos included. 
We want to come back to this problem in future. (See also Appendix \ref{AppB}.)

\section{Hawking radiation with a chemical potential} \label{AppB}
Here we briefly explain why the chemical potential modifies the spectrum of Hawking radiation.
We start from the following action in the Schwartzschild black hole geometry.
%
\al{
S&=
\int \sqrt{-g}\,d^4x 
\paren{
\bar{\psi}
i\not\!\partial
\psi
+
{C\over M_*^4}
g^{\mu\nu}
\partial_\mu 
\paren{\mathcal{R}_{\alpha\beta\gamma\delta}\mathcal{R}^{\alpha\beta\gamma\delta}} 
\bar{\psi}
\not\!\partial
\psi
}
\nn
&=
\int \sqrt{-g}\,d^4x 
\,
\bar{\psi}
\paren{
i\partial_\mu - {C\over M_*^4} 
\partial_\mu
\paren{\mathcal{R}_{\alpha\beta\gamma\delta}\mathcal{R}^{\alpha\beta\gamma\delta}} 
}
\gamma^\mu
\psi
.
}
The action has the same form as 
\al{
\int \sqrt{-g}\,d^4x 
\,
\bar{\psi}
\paren{
i\partial_\mu - e A_\mu}
\gamma^\mu
\psi,
}
by identifying $C\,\partial_\mu\paren{\mathcal{R}_{\alpha\beta\gamma\delta}\mathcal{R}^{\alpha\beta\gamma\delta}}/M_*^4$ with the gauge potential $A_\mu$.
Thus the coupling to the background gravity is nothing but the pure gauge configuration (but it cannot be gauged away as  discussed in the previous Appendix).
It is  now instructive to briefly sketch the derivation of Hawking radiation in the case of the charged black hole following Refs.~\cite{Iso:2007kt,Iso:2007hd}.
At the outer event horizon, the action of the charged fermion becomes
\al{\label{Eq:charged black hole}
S&\simeq
\int dt dr_* 
\paren{
\bar{\psi}
\paren{
i\partial_t - e A_t
} 
\gamma^t
\psi
-
\bar{\psi}
i\partial_{r_*}
\gamma^{r_*}
\psi
},
&
A_t=-{eQ\over r} .
}
Here we omit the contribution from angular components for simplicity, and $Q$ is the charge of the black hole.
In order to obtain the outgoing flow of the energy  at the future infinity, 
we usually impose the ingoing boundary condition for the current at the horizon.
But since the above scalar potential diverges $A_U \propto A_t/U$ at the horizon $U=0$  in the Kruskal coordinate, we need 
 to take the gauge in which $A_t=0$  at the horizon (see discussions e.g. \cite{Iso:2007kt,Iso:2007hd})
\al{\label{Eq:gauge tr}
&
A_t' = A_t + \partial_t \Lambda,
&
\Lambda = {eQ\over r_H} t
.
}
This shifts the energy level of the charged particles and  the spectrum of the Hawking radiation is obtained by replacing
$\omega$ by $\omega -\mu=eQ/r_H$.  Thus $eQ/r_H$  indeed plays the role of the chemical potential in the thermal radiation.

The same procedure is applicable to the current setup.
In our case, the ``gauge transformation" to regularize the action at the horizon in the Kruskal coordinate  is given by 
\al{
\psi
\to
\exp\paren{i{C\over M_*^4}\partial_t\paren{\mathcal{R}_{\mu\nu\rho\sigma}\mathcal{R}^{\mu\nu\rho\sigma}}\bigg|_{r=r_H}t}
\psi.
}
Thus 
\al{
\mu=
{C\over M_*^4}
\partial_t
\paren{\mathcal{R}_{\alpha\beta\gamma\delta}\mathcal{R}^{\alpha\beta\gamma\delta}} 
\bigg|_{r=r_H}
}
becomes the chemical potential to describe  the thermal radiation from  the PBH.

\section{Analytical approximation}\label{App:analytic}
In this appendix, we present an analytical approximation of the function $g_n(X)$ which is defined in Eq. (\ref{gnX}) and 
appears in the calculation of the asymmetry. We investigate the large $n$ behavior of the integral.
For large $X$, it can be expanded with respect to $1/X$ as
\al{
g_n(X)
\simeq
{40g_L\over7\paren{(2g_L+g_{\rm other})\pi^2}}{1\over X^{2n+1}}
+
{40g_L(46g_L-7g_{\rm other})\over49(2g_L+g_{\rm other})^2\pi^4}\paren{1\over X^{2n+1}}^3
+ \cdots
}
For $X \rightarrow \infty$, 
the first term gives a good approximation for $g_n(X)$ and we have an approximated formula for the integral; 
\al{
&f_n(X_{\rm min}, X_{\rm max}) =\int^{X_{\rm max}}_{X_{\rm min}} dX g_n(X)
\simeq
-{20\over7\pi^2}{g_L\over2g_L+\dfrac{1}{3}g_{\rm other}}{1\over n}{1\over X_{\rm min}^{2n}},
}
for $1\leq  X_{\rm min}\leq X_{\rm max}$. 
Then, from Eq.(\ref{deltaNLn}), we have
 $\delta N_L \propto M_{\rm cr}^2 {X_{\rm min}^{-2n}}$. In this region ($X>1$), $T_H>\mu$ and $X_{\rm min}$ is given by $X_{\rm min}=(M_P^2/M_{\rm cr}M_*)^2$. By using 
 the result for the critical mass $M_{\rm cr}$ in (\ref{ninf-criticalmass}), it turns out that $\delta N_L$ is independent of the mass scale $M_*$.
 Since the entropy $s$ is determined by the life time of the PBH, namely $M_{\rm PBH}$, and independent of $M_*$, $n_L/s$ also becomes independent of $M_*$.
 The region $1<X_{\rm min}$ is given by the region
 where the line of $M_{\rm PBH}=M_{\rm cr}$ is below the line of $T_H(M_{\rm PBH})=M_*$, and 
  corresponds to the region with $M_*\lesssim 10^{-3}M_P$ in the left panel of Fig.~\ref{Fig:asymmetry}. 
This is the reason why the generated asymmetry is constant as a function of $M_*$ for $M_*\lesssim 10^{-3}M_P$.
But we notice that the behavior is given by our prescription to cut the integral at $X_{\rm min}$ where the typical energy of the Hawking radiation becomes 
higher than $M_*$ and the present investigations become questionable.
Though it is beyond our approximation to estimate $\delta N_L$ for $X\lesssim X_{\rm min}$, 
 larger asymmetry may be produced in this region. 
In this sense, the estimation in this paper would give a conservative value for the lepton asymmetry.

\begin{figure}
\begin{center}
\includegraphics[width=.49\textwidth]{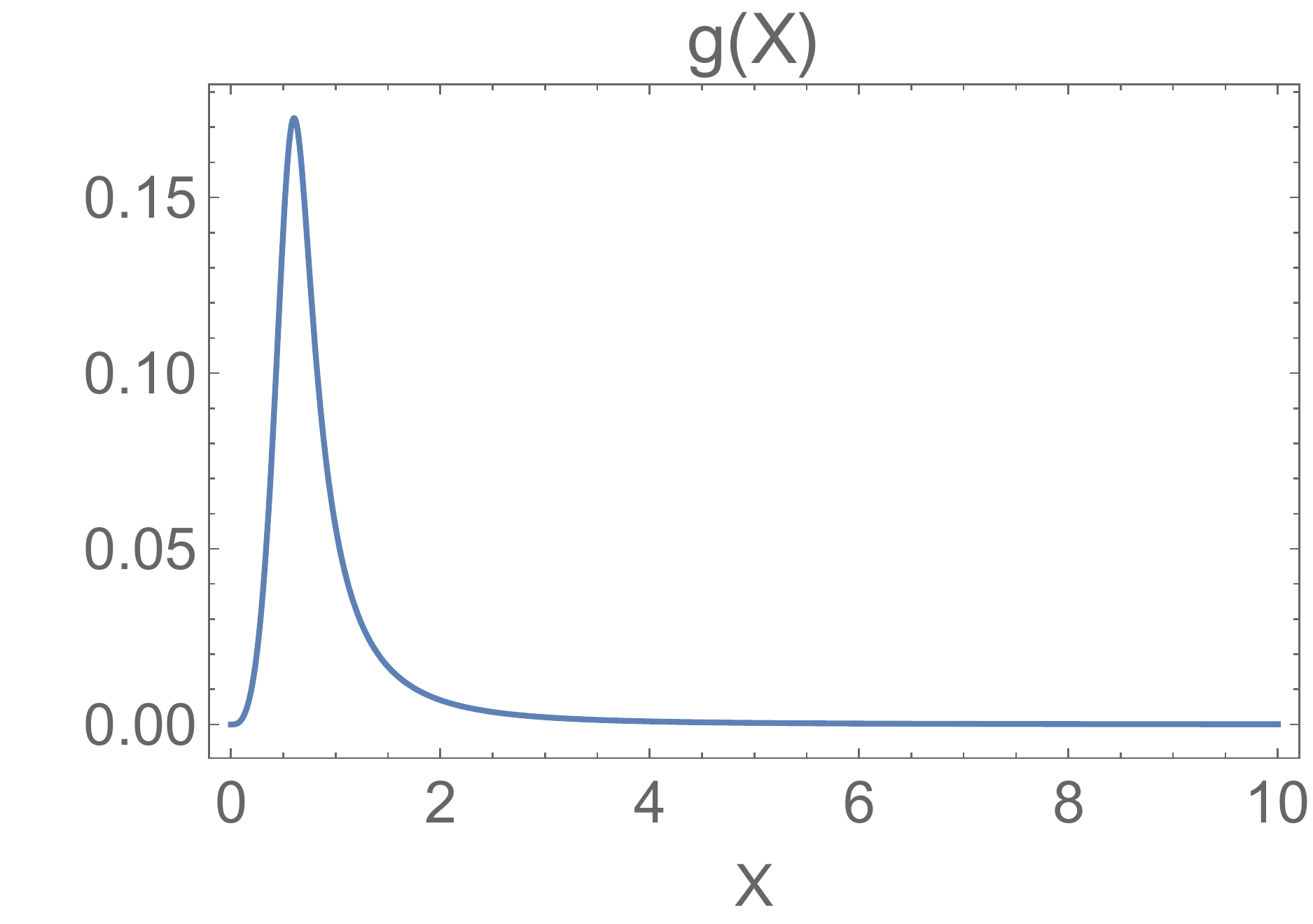}
\hfill
\includegraphics[width=.49\textwidth]{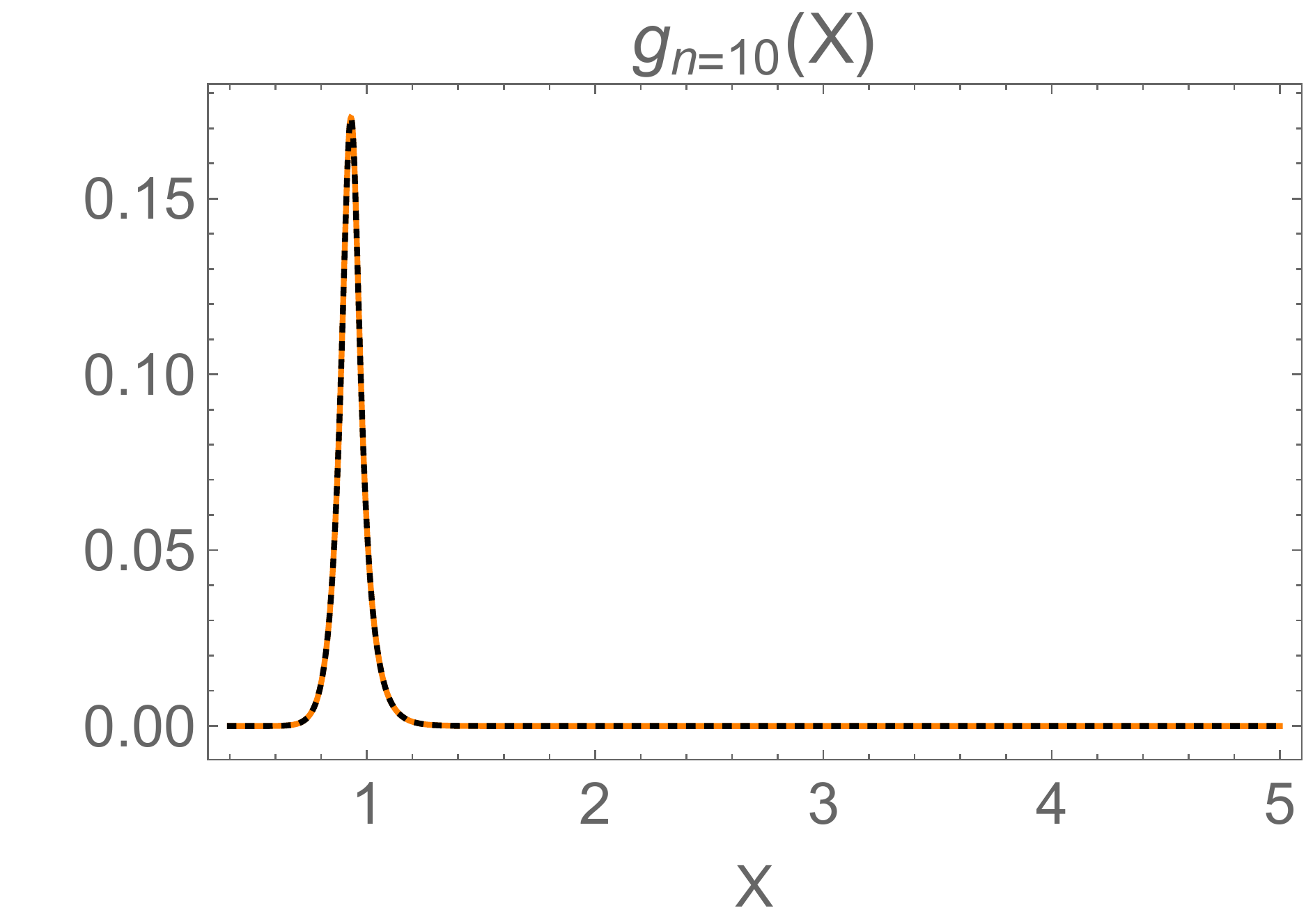}
\hfill\mbox{}
\end{center}
\caption{
The solid orange and dashed black lines correspond to $g(X)$ and $g_{\rm app}(X)$, respectively.
One can see that $g_{\rm app}(X)$ is good approximation for all the region of $X$, and the integrand has strong peak around $X\sim1$.
Note that the position of the peak for $n=10$ is closer to $X=1$ than for $n=1.$
}
\label{Fig:integrand approximation}
\end{figure}

On the other hand, in the region $X\sim0$, we have
\al{\label{Eq:integrand approximation}
g_{\rm app}(X)&=
-{40g_L X^{2n+1}\paren{1+\pi^2 \paren{X^{2n+1}}^2}\over 30 g_L+60g_L\pi^2\paren{X^{2n+1}}^2+7(2g_L+g_{\rm other})\pi^4\paren{X^{2n+1}}^4},
\nn
&=
-{4\over3}{C X^{2n+1}\paren{1+\pi^2 \paren{X^{2n+1}}^2}\over C+2C\pi^2\paren{X^{2n+1}}^2+\paren{X^{2n+1}}^4}, 
}
where  
$C:=30g_L/(7\pi^4(2g_L+g_{\rm other}))$.
Surprisingly, as shown in Fig.~\ref{Fig:integrand approximation}, this expression gives a good approximation even for $X>1$ as well as for $X\sim0$.
%
Since Eq.~\eqref{Eq:integrand approximation} is a rational function of $X$,  its integral can be performed analytically, and we have
{\footnotesize
\al{\label{Eq:integral approximation}
&f_{\rm app}(X_{\rm min},X_0):=\int^{X_0}_{X_{\rm min}} g_{\rm app}(X)
%
=
   -{1\over3}
   {\sqrt{C}\over\sqrt{C \pi^4-1}} \Bigg[{1\over n+1}
   \pi ^2 \bigg\{
   \nn
   &
   -H_n(X_{\rm min},X_{\rm min})
   G_{n-}(X_0)
   F_{--}(X_0)
   {X_0}^{6 n+4}
   +
   H_n(X_0,X_0)
   G_{n-}(X_{\rm min})
   F_{--}(X_{\rm min})
   {X_{\rm min}}^{6 n+4}
   \nn
   & 
   +
   H_n(X_{\rm min},X_{\rm min})
   G_{n+}(X_0)
   F_{-+}(X_0)
   {X_0}^{6 n+4}
   -
   H_n(X_0,X_0)
   G_{n+}(X_{\rm min})
   F_{-+}(X_{\rm min})
   {X_{\rm min}}^{6 n+4} 
   \bigg\}
   \nn
   &
   +{1\over n}\bigg\{
   %
   H_n(X_{\rm min},X_{\rm min})
   G_{n-}(X_0)
   F_{+-}(X_0)
   {X_0}^{2n+2}
   - 
   H_n(X_0,X_0)
   G_{n-}(X_{\rm min})
   F_{+-}(X_{\rm min})
   {X_{\rm min}}^{2n+2}
   \nn
   &
   -
   H_n(X_{\rm min},X_{\rm min})
   G_{n+}(X_0)
   F_{++}(X_0)
   {X_0}^{2 n+2}
   +
   H_n(X_0,X_0)
   G_{n+}(X_{\rm min})
   F_{++}(X_{\rm min})
   {X_{\rm min}}^{2n+2} 
   \bigg\}
   \Bigg]
   \nn
   &
   \times
   \sqbr{
   (X_0  X_{\rm min})^{4 n+2}(2C\pi^2+X_0^{4n+2})(2C\pi^2+X_{\rm min}^{4n+2})
   +C (
   H_n(X_0,X_0)
   +
   H_n(X_{\rm min}, X_{\rm min})
   -
   C
     )
   }^{-1}.
}
}
Here we have defined
\al{
&
F_{\pm\pm}(Y):=
_2F_1\left(1,1;{2n+(1/2)\over2n+1}\pm{1\over2};\frac{1}{\left(\pi ^2\pm\frac{\sqrt{C \pi^4-1}}{\sqrt{C}}\right) Y^{4 n+2}+1}\right),
   \nn
&
G_{n\pm}(Y):=Y^{4n+2}\pm\sqrt{C}\sqrt{C\pi^4-1}+C\pi^2,
\nn
& H_n(Y,Z):=2C\pi^2 Y^{4n+2}+Z^{8n+4}+C,
}
where $_2F_1$ is the hypergeometric function.
In the numerical calculation for drawing the figures, we used the expression of Eq.~\eqref{Eq:integral approximation}.

Finally let us examine the large $n$ behavior of $f_{\rm app}(X_0,X_{\rm min})$.
We concentrate on the region $X_0>1$ for simplicity.
The behavior of $X_{\rm min}$ is given by
\al{&
X_{\rm min}\to 1-{1\over8n}\paren{10\log 2+\log 3+4\log\pi},
}
and thus
\al{
X_{\rm min}^{2n+2}\to \frac{1}{4\pi\times{3}^{1/4}\sqrt{2}}:=D\simeq0.043.
}
%
Then the hypergeometric function is approximated for large $n$ as
\al{
F_{\pm\pm}(Y)&\to
 _2F_1\left(1,1;1\pm{1\over2};\frac{1}{\left(\pi ^2\pm\frac{\sqrt{C \pi^4-1}}{\sqrt{C}}\right) Y^{4 n+2}+1}\right)
\nn
\to &
\begin{cases}
			1	&	
				\text{for $Y=X_0$,}\\ \\
			_2F_1\left(1,1;1\pm{1\over2};\frac{1}{\left(\pi ^2\pm\frac{\sqrt{C \pi^4-1}}{\sqrt{C}}\right) D^{-2}+1}\right)	&	
				\text{for $Y=X_{\rm min}$.}	
		\end{cases}
}
Note that
\al{&
_2F_1\left(1,1;{3\over2};x\right)={\arcsin(\sqrt{x})\over\sqrt{x-x^2}},
&&
_2F_1\left(1,1;{1\over2};x\right)={1\over1-x}+{\sqrt{x}\arcsin(\sqrt{x})\over(1-x)^{3/2}}.
}

Then, by picking up the leading power of $X_0$, $f_{\rm app}$ becomes
\al{
f_{\rm app}&\to
-{1\over3}
{1\over n} \sqrt{C\over C\pi^4-1}
\nn
&
\times
X_0^{8n+4}
\big[
\pi^2 D^3\paren{G_{n-}(X_{\rm min})F_{n--}(X_{\rm min})-G_{n+}(X_{\rm min})F_{n-+}}(X_{\rm min})
\nn
&
+D\paren{-G_{n-}(X_{\rm min})F_{n+-}(X_{\rm min})+G_{n+}(X_{\rm min})F_{n++}(X_{\rm min})}
\big]
\nn
&
\times{1\over X_0^{8n+4}\sqbr{D^2\paren{2C\pi^2+D^2}+C}}
\nn
&=
{1\over n} \sqrt{C\over C\pi^4-1}{1\over\sqbr{D^2\paren{2C\pi^2+D^2}+C}}
\nn
&
\times
\big[
\pi^2 D^3\paren{G_{n-}(X_{\rm min})F_{n--}(X_{\rm min})-G_{n+}(X_{\rm min})F_{n-+}}(X_{\rm min})
\nn
&
+D\paren{-G_{n-}(X_{\rm min})F_{n+-}(X_{\rm min})+G_{n+}(X_{\rm min})F_{n++}(X_{\rm min})}
\big],
}
from which it is concluded that $f_{\rm app}\sim 1/n$, and vanishes for $n\to\infty$.

\subsection*{Acknowledgement}
We thank Koichi Hamaguchi, Ryuichiro Kitano, Kazunori Kohri, Sujoy Kumar Modak, Pasquale Serpico, Kengo Shimada and Hiroshi Umetsu for helpful discussions on various aspects 
of the baryogenesis and the PBH. Especially Hamaguchi and Kitano asked critical questions which stimulated our further investigations. 
We also acknowledge fruitful conversations with the participants of 
KEK summer camp held in Azumino, Nagano during August 3 and 8, 2016. 
The work of YH is supported by the Grant-in-Aid for Japan Society for the Promotion of Science Fellows, No. 16J06151.
This work (SI) is supported by the Grant-in-Aid for Scientific research from the Ministry of Education, Science,
Sports, and Culture, Japan, Nos. 23540329. 

\bibliographystyle{TitleAndArxiv}
\bibliography{refs}

\end{document}